\documentclass[aip,jcp,reprint]{revtex4-1}

\usepackage{graphicx}
\usepackage{subfigure}
\usepackage{amsmath, amsthm}
\usepackage{color}
\usepackage{tcolorbox}
\usepackage{lipsum}

\usepackage{graphicx}
\usepackage{subfigure}
\usepackage{amsmath, amsthm}
\usepackage{color}
\usepackage{tcolorbox}
\DeclareMathOperator\erf{erf}

\begin{document}

\title{First-Order ``Hyper-selective'' Binding Transition of Multivalent Particles Under Force}
\author{Tine Curk}
	\affiliation{Materials Science and Engineering, Northwestern University, Evanston, Illinois 60208, United States}
\author{Nicholas B. Tito}
	\affiliation{Department of Applied Physics, Eindhoven University of Technology, PO Box 513, 5600 MB, Eindhoven, The Netherlands}
	\affiliation{Institute for Complex Molecular Systems, Eindhoven University of Technology, PO Box 513, 5600 MB, Eindhoven, The Netherlands}
	\email{nicholas.b.tito@gmail.com}
	
\date{\today}
\begin{abstract}

Multivalent particles bind to targets via many independent ligand-receptor bonding interactions. This microscopic design spans length scales in both synthetic and biological systems. Classic examples include interactions between cells, virus binding, synthetic ligand-coated micrometer-scale vesicles or smaller nano-particles, functionalised polymers, and toxins. Equilibrium multivalent binding is a continuous yet super-selective transition with respect to the number of ligands and receptors involved in the interaction. Increasing the ligand or receptor density on the two particles leads to sharp growth in the number of bound particles at equilibrium.

Here we present a theory and Monte Carlo simulations to show that applying mechanical force to multivalent particles causes their adsorption/desorption isotherm on a surface to become sharper and more selective, with respect to variation in the number of ligands and receptors on the two objects. When the force is only applied to particles bound to the surface by one or more ligands, then the transition can become \emph{infinitely} sharp and first-order---a new binding regime which we term ``hyper-selective''. Force may be imposed by, e.g. flow of solvent around the particles, a magnetic field, chemical gradients, or triggered uncoiling of inert oligomers/polymers tethered to the particles to provide a steric repulsion to the surface. This physical principle is a step towards ``all or nothing'' binding selectivity in the design of multivalent constructs. 
\end{abstract}

\maketitle

\section{Introduction}

Multivalent particles are microscopic objects that interact with each other by many independent bonding units, often called ``ligands'' and ``receptors''.\cite{Bell:1978hj,Bell:1984wv,Sulzer:1996dd,Huskens:2004je,MartinezVeracoechea:2011kn,Varilly:2012gl,AngiolettiUberti:2013iu,MartinezVeracoechea:2013ih} Multivalent interactions are a potent binding motif due to their \emph{super-selectivity}\cite{MartinezVeracoechea:2011kn, Varilly:2012gl}, wherein the number of bound multivalent particles to a target increases sharply with the density of receptors on the target. Living organisms have evolved to depend on multivalent binding paradigms in some of their most delicate and mission-critical pathways, e.g. chemical communication at and between cell surfaces, interactions between biomolecular complexes and cells, viral/bacterial adhesion, and (extra-)cellular machinery.

Instances of multivalent interactions span from small to large length scales. Structures that exhibit multivalent interaction at small length scales include functionalised (bio-)polymers \cite{Varner:2015dh,Dubacheva:2014ka,Dubacheva:2015hca,Dubacheva:2019gq}, nanoparticles, biological toxins, and viruses.\cite{Mammen:1998im,Hlavacek:2002dv,Vonnemann:2015im,Xu:2016kk,Liese:2018eu,DiIorio:2019jy,Muller:2019dk}  At larger length scales, cells in living organisms have a multitude of different kinds of receptors on their surfaces/membranes, which serve as points of communication with the outside world. Interactions between cells are often multivalent.\cite{Bell:1978hj,Macken:1982ia,Bell:1984wv,Perelson:1980ds,Sulzer:1996dd,Sulzer:1997gq,Hlavacek:2002dv,Chen:2003fe,Evans:2007gl, Hong:2007ek,Carlson:2007hv,Shimobayashi:2015fz,Xu:2016kk,Weikl:2016bj,Curk:2017cj,Amjad:2017io,AngiolettiUberti:2017iv,DiMichele:2018dm, Vahey:2019gq} On the synthetic side, classic multivalent constructs include ligand-coated colloids and vesicles, often employing DNA in order to finely tune their interactions.\cite{Mirkin:1996em,Biancaniello:2005ie,Rogers:2011jp,Varilly:2012gl,AngiolettiUberti:2012gi,AngiolettiUberti:2013iu,Wu:2013jg,Stoffelen:2014gr,MejiaAriza:2014jm,AngiolettiUberti:2014kl,Li:2015bm,Wang:2015ep,Curk:2018jb,Srinivasan:2013ez,Grindy:2016jf,Bachmann:2016bb,Newton:2015dq,Theodorakis:2015it,Myers:2016ca,vanderMeulen:2015jb,Newton:2017fp,AngiolettiUberti:2016dd,Mbanga:2016ej,Stoffelen:2015er,DiMichele:2016iu,Zhang:2017kw,Halverson:2016cq,Lanfranco:2019cl,Post:2019hk} Mixtures of different kinds of multivalent particles can be designed to sequentially self-assemble, or to exhibit remarkably selective surface adsorption.\cite{Tito:2016hh,DiMichele:2016iu,Zhang:2017kw,Halverson:2016cq,Tito:2019fk} However, the kinetics of multivalent interactions play a strong role in whether the system reaches equilibrium, or a non-trivial kinetic steady-state (particularly for strong-binding ligands with long lifetimes). \cite{Block:2016fi,Weikl:2016bj,Bachmann:2016dm,Vijaykumar:2018jq,Licata:2008kp,Newton:2015dq,Newton:2017fp,Lanfranco:2019cl}

Binding of multivalent particles is a continuous transition at equilibrium. There are both enthalpic and entropic contributions to their adhesion strength. The enthalpic contribution, intuitively, arises from the bonding between the ligands and receptors. More bonds mean a larger, more negative, and more favourable enthalpic contribution to the binding free energy.

The entropic contribution is less obvious. Firstly, ligands and receptors must lose local configurational entropy in order to make a bond. This leads the ``effective'' ligand/receptor bond strength to often be lower than what is observed between the two structures in, for example, free solution.\cite{Varilly:2012gl,MartinezVeracoechea:2013ih} Secondly, there is a \emph{favourable} entropic binding contribution to the number of possible binding permutations that the ligands and receptors may explore. If the ligands and receptors are short and spaced far apart, then this entropy reflects the fact that each bond can be independently bound or unbound. If the ligands and receptors are long and flexible, then an additional source of entropy is the number of binding partners that each entity may have, much like making connections on a telephone switchboard.\cite{Tito:2016hh}

The permutation entropy becomes larger and more favourable when there are more ligands and receptors on the two multivalent structures. Thus, the binding free energy $\Delta G$ grows more negative, and the binding probability grows exponentially larger (since this depends on $\exp{(-\Delta G/RT)}$). This rapid growth in the binding probability with the number of ligands and receptors on the two objects is referred to as super-selectivity. It is fundamentally an entropic effect. For example, \emph{monovalent} binders can never exhibit super-selective binding, since they lack the permutation contribution to their individual binding free energy. Their bonding strength may only be modulated by the enthalpy of their (single) bond.

This study examines in detail how the microscopic thermodynamics of multivalent binding change when mechanical force is applied to the particles. In the biological arena, objects bound to cell surfaces are often exposed to flow (e.g. in blood vessels) or other sources of force in the extracellular matrix. Force, via magnetic fields or electric charges, is also a convenient tool for manipulating synthetic multivalent systems. The \emph{response} of a multivalent object to force, e.g. using atomic force microscopy (AFM)\cite{Auletta:2004fs, Erdmann:2008jf, GomezCasado:2011gh, Bacharouche:2015em} or single-molecule force spectroscopy \cite{Evans:2007gl}, can also serve as a probe for the strength and type of interactions it has with its target.

\begin{figure}
	\centering
	\includegraphics[width= 0.50\textwidth]{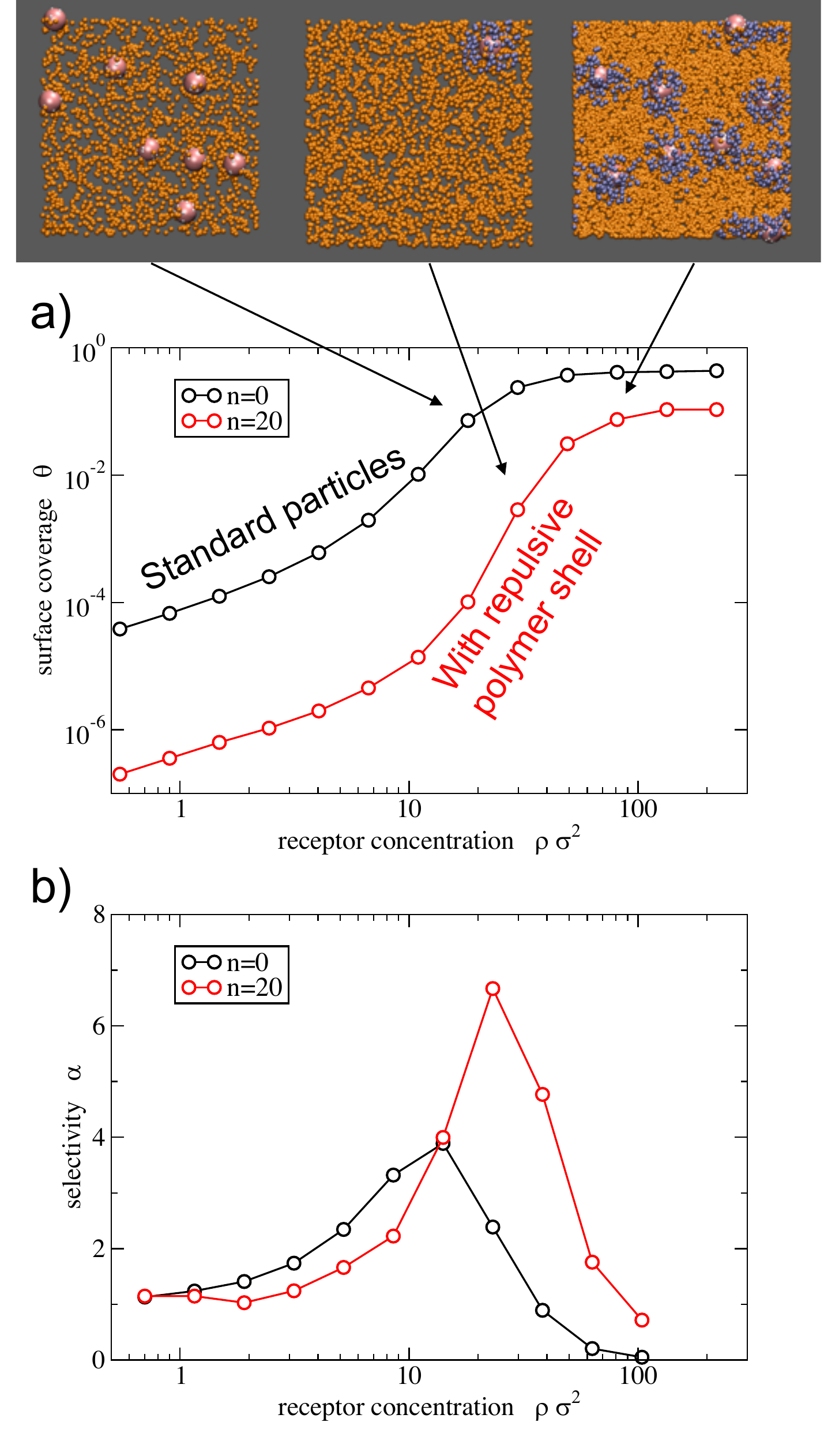}
	\caption{Grand canonical Monte Carlo simulations of a ligand coated nanoparticle showing the addition of inert polymers increases the binding selectivity. The snapshots correspond to the points indicated by the arrows and demonstrate the super--selective adsorption behaviour: changing the receptor concentration by a factor 2.7 induces a 9--fold increase in the adsorbed particle density. The receptors and ligands are coloured orange, the particles are pink and the inert polymer are blue/silver.  a) the surface density of adsorbed particles as function of the receptor concentration, the black curve in corresponds to the reference system and the red curve corresponds to the particles coated with $n=20$ inert polymers.  b) the selectivity, {\em i.e.}, the slope of the curves in a). The model and algorithm are explained in Ref. \citenum{Curk:2017cj}. Parameters: ligand--receptor bond energy $\epsilon=3k_{\rm B}T$, number of ligands $k=20$, inert polymer length length $n_{\rm ip}=5$ blobs, the imposed chemical potential maintains the particle concentration in bulk solution at $\rho_p=0.001/\sigma^3$, with $\sigma$ the particle hard--core diameter. Lateral system size $L_x=L_y=10\sigma$. $\theta$ is measured in units of $\sigma^2$.}
	\label{fig:TineSims}
\end{figure} 

To motivate our work with a concrete example, consider the Monte Carlo simulation results of multivalent particle binding in Figure \ref{fig:TineSims}. These simulations comprise explicit spherical particles (pink) coated with bead-spring ligands (orange), interacting with a flat surface with explicit receptors (also orange). Long inert polymers (blue) can also be tethered to the surfaces of the multivalent particles; the entropic and excluded-volume repulsion between these polymers and the surface effectively impose a normal force on bound particles when close to the surface. Coating particles with inert polymers is an example of an equilibrium system that exhibits a tuneable effective normal force (albeit the coating also provides a lateral repulsion between the particles). Details of the Monte Carlo model can be found in Ref. \citenum{Curk:2017cj}.

Figure \ref{fig:TineSims} shows simulated adsorption profiles for multivalent particles with 20 ligands per particle. The black curve shows the adsorption profile for the particles with no inert polymers. Adding $n=20$ inert polymers grafted uniformly at random on the particle's surface (red curve) serves to increase the sharpness of the adsorption profiles characterised by the slope or selectivity $\alpha$. The receptor concentration where the inflection point occurs also increases. These trends are noted by Wang and Dormidontova in simulations of multivalent particles with a bimodal distribution of ligand lengths.\cite{Wang:2012cs}

This work now develops a quantitative theoretical handle on how applied force affects the selectivity of multivalent binding. We also elucidate under which conditions the binding actually becomes \emph{first-order} and \emph{discontinuous}. This is a new multivalent binding regime which we refer to as the ``hyper-selective'' regime.  The transition is characterised by a discontinuity in the equilibrium free energy per particle as a function of the number of receptors on the target surface.

To start, a model for the equilibrium response of a bound multivalent particle to a pulling force is derived. Attention is restricted to the simple scenario of multivalent particles bound to a substrate with mobile receptors at a fixed non-depleting concentration. We then consider what happens when a constant force is applied to the particles normal to the surface. A crucial distinction is made between two cases: first, when \emph{both} the unbound and bound particles are exposed to the force field; and second, when the force field only affects the \emph{bound} particles. From our theory, we extract a clear microphysical understanding of what leads multivalent particles to exhibit hyper-selective binding, and how the transition depends on the design and concentration of the particles. In the conclusions we outline equilibrium and non-equilbrium strategies for realising enhanced super-selective and hyper-selective binding.

\section{Model for Multivalent Force-Extension Response}

\begin{figure*}
	\centering
	\includegraphics[width= 0.90\textwidth]{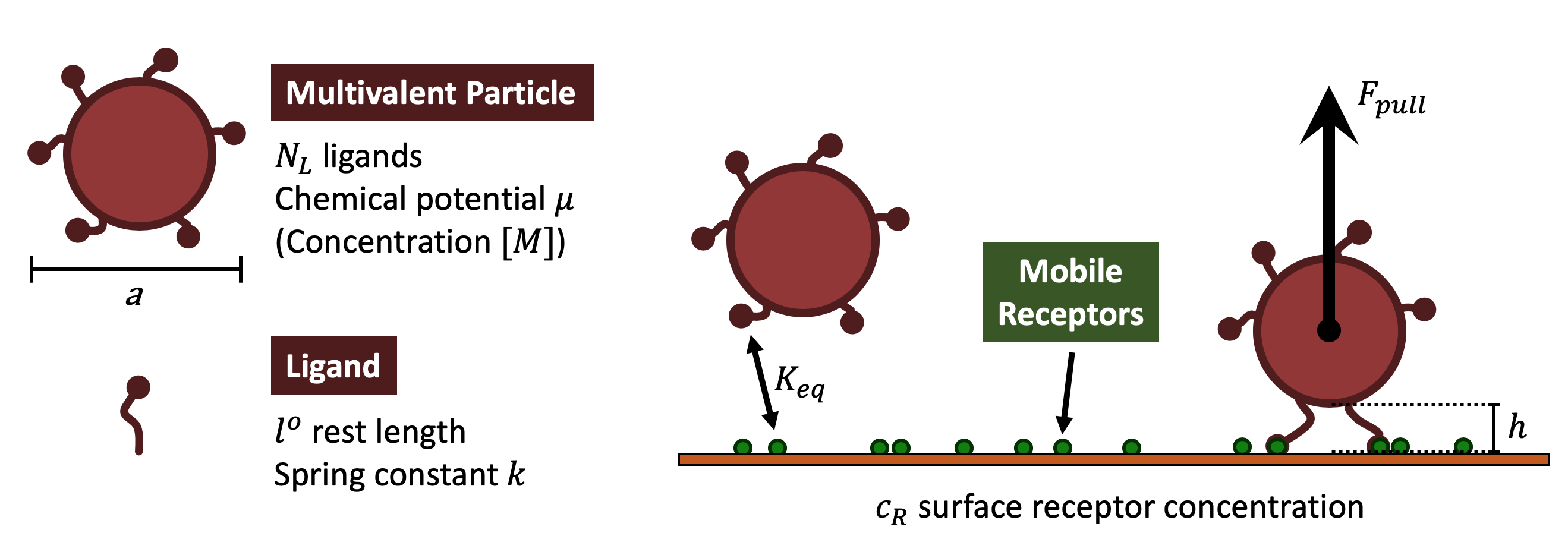}
	\caption{Illustration of ingredients in the multivalent binding model. Quantities are defined in text.}
	\label{fig:Cartoon}
\end{figure*}  

Consider a multivalent particle with ligands that interact with \emph{mobile} receptors on an adjacent flat surface. Components of the model are illustrated in Figure \ref{fig:Cartoon}. Let the quantity $N_L$ define the number of ligands on a particle that are within reach of the receptor surface. The density of receptors on the surface is $c_R$ in units of moles of receptors per $b^2$, where ``$b$'' is the distance unit of the model. We will assume that the receptors cannot be depleted, i.e. they come from a reservoir at fixed surface concentration $c_R$. Energy units are in terms of $RT$, where $R$ is the ideal gas constant and $T$ is absolute temperature.

The ligands are treated as Hookian springs with a spring constant $k$ (in units of energy per squared distance) and rest length $l^\circ$. The receptors are considered to be points on the substrate. The ligand/receptor association constant in free solution is denoted $K_{\text{eq}}$ (in units of $b^3 / $mol).

The theory developed in Appendix \ref{app:Theory} uses equilibrium statistical mechanics to predict the quasi-equilibrium ``force versus extension'' curve for a multivalent particle: that is, how the restoring force $F(h)$ depends on the particle height $h$.  The quasi--equilibrium regime is obtained when the rate at which force is being applied on the multivalent particles is vanishingly small. As a result, the system is quasi-static and attention is restricted only to the quasi-equilibrium thermodynamics.

The starting point for the model is the binding free energy per ligand.  This expression, derived in detail in Ref. \citenum{Curk:2017cj}, takes an equilibrium ensemble average over: a Poisson distribution of mobile receptors within the surface contact area of a multivalent particle; and over all possible ligand/receptor binding permutations.  The resulting expression has three contributions:
\begin{equation}
	\frac{\Delta G^b_{\text{lig}} (h)}{RT} = \frac{k (h - l^\circ)^2}{2 RT} - \ln{\left(\frac{c_R K_{\text{eq}}}{h}\right)} + \frac{\Delta G^b_{\text{lig},\text{cnf}} (h)}{RT}.
	\label{eqn:FESingleLigandMainText}
\end{equation}
The first term accounts for the ``stretch energy'' of the ligands from their ideal lengths $l^\circ$. The second term accounts for the strength of the ligand/receptor bond (via $K_{\text{eq}}$), and the effective molarity $c_R / h$ of receptor binding partners. The third term is the configurational free energy of the ligand when it is in the bound state, and confined within the region $h$ between the multivalent particle and the substrate. 

When a ligand is unbound, then the only contribution to its free energy is its configurational entropy within the gap $h$:
\begin{equation}
	\frac{\Delta G^{ub}_{\text{lig}} (h)}{RT} = \frac{\Delta G^{ub}_{\text{lig},\text{cnf}} (h)}{RT}.
\end{equation}
The precise forms of the configurational free energies $\Delta G^{ub}_{\text{lig},\text{cnf}} (h)$ and $\Delta G^{b}_{\text{lig},\text{cnf}} (h)$ are derived and presented in Appendix \ref{app:Theory}.

\begin{figure}
	\centering
	\includegraphics[width= 0.48\textwidth]{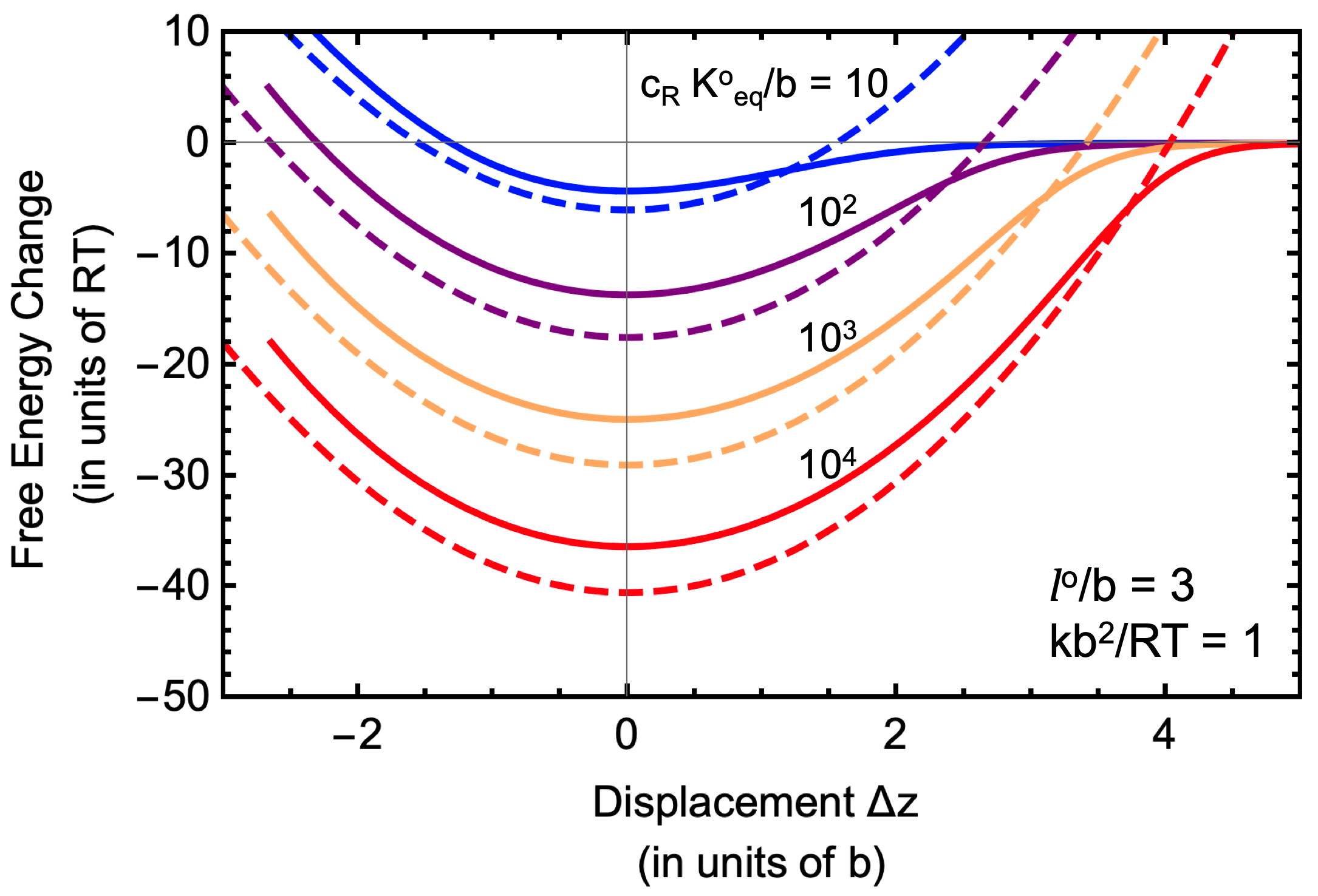}
	\caption{Plots of the multivalent binding free energy, Eq. \ref{eqn:FEMainText}, as a function of relative separation distance $\Delta z$ between the receptor surface and multivalent particle exterior. Results are shown for four choices of overall binding strength, $c_R K_{\text{eq}}/b$, given that the multivalent particle has $N_L = 5$ ligands each with an equilibrium length $l^\circ/b = 3$ and stiffness $k b^2/RT = 1$. Solid lines are numerical calculations using Eq. \ref{eqn:FEMainText}, and dashed lines are the approximation given by Eq. \ref{eqn:FEMainTextApprox}.}
	\label{fig:FE}
\end{figure}  

\begin{figure*}
	\centering
	\includegraphics[width= 0.9\textwidth]{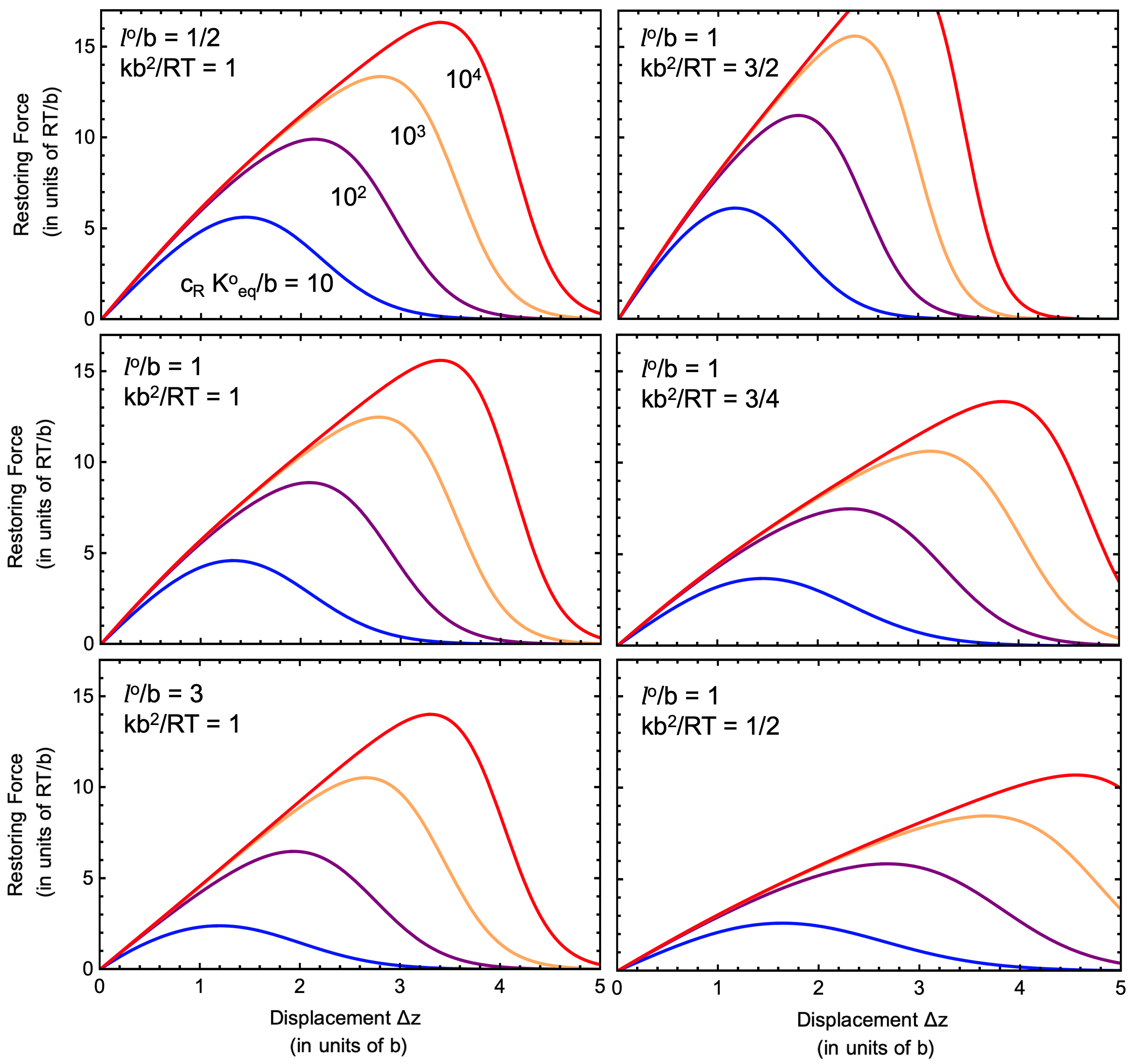}
	\caption{Plots of the restoring force (Eq. \ref{eqn:ForceReal}) as a function of relative separation distance $\Delta z$ between the receptor surface and multivalent particle exterior. Each panel shows results for four choices of the ligand-receptor binding strength, $c_R K_{\text{eq}}/b$, assuming the multivalent particle has $N_L = 5$ ligands. The ligand rest lengths $l^\circ$ and stiffness $k b^2 / RT$ are indicated within the panels.}
	\label{fig:ForceCurves}
\end{figure*} 

Given the bound- and unbound-state ligand free energies, the full binding free energy for the multivalent particle is
\begin{equation}
	\frac{\Delta G_{\text{bind}}(h)}{RT} = - N_L \ln{\left(e^{-\Delta G^{ub}_{\text{lig}} (h)/RT} + e^{-\Delta G^b_{\text{lig}} (h)/RT}\right)},
	\label{eqn:FEMainText}
\end{equation}
representing the fact that each of the $N_L$ ligands on the particle can be independently bound or unbound. The height coordinate $h$ corresponding to the minimum of $\Delta G_{\text{bind}}(h)$ is defined to be $h_{\text{min}}$.

Figure \ref{fig:FE} presents plots of the multivalent binding free energy, $\Delta G_{\text{bind}}(h)$, as a function of the relative separation distance $\Delta z \equiv h - h_{\text{min}}$ between the receptor surface and the particle exterior. At values of $h < h_{\text{min}}$, the free energy grows more unfavourable due to the entropy loss associated with ligand confinement, contained in $\Delta G^{ub}_{\text{lig},\text{cnf}} (h)$ and $\Delta G^{b}_{\text{lig},\text{cnf}} (h)$. For values of $h > h_{\text{min}}$ the free energy again grows more unfavourable due to: a decrease in the average number of bound ligands, and the stretch free energy associated with the ligands that are bound (i.e. the first term in Eq. \ref{eqn:FESingleLigandMainText}).

In Appendix \ref{app:Approximations}, we derive a simple approximation for the multivalent binding free energy profiles when the ligands are strong-binding (i.e. large $c_R K_{\text{eq}}$):
\begin{equation}
	\frac{\Delta G_{\text{bind}}(h)}{RT} \approx N_L \left[\frac{k (h - l^\circ)^2}{2 RT} - \ln{\left(\frac{c_R K_{\text{eq}}}{l^\circ}\right)}\right],
	\label{eqn:FEMainTextApprox}
\end{equation}
Calculations using this equation are shown in Figure \ref{fig:FE} as dashed lines, defining the relative displacement $\Delta z = h - l^\circ$ (noting that $h_{\text{min}} = l^\circ$ in Eq. \ref{eqn:FEMainTextApprox}). We see that this form well captures the parabolic curvature of the exact free energy profiles, as well as the scaling of their minimum values with $c_R K_{\text{eq}} / b$.

The restoring force is calculated by taking the gradient of the binding free energy, $d \Delta G_{\text{bind}}(h) / d h$ (ignoring the typical negative sign so that our force values are positive). This is
\begin{align}
	F(h) = &N_L \left[P_{b,1}(h)  \left(k (h - l^\circ) + \frac{RT}{h} + \left.\frac{d \Delta G^b_{\text{lig},\text{cnf}}(h)}{d h}\right\vert_{h}\right) \right. \nonumber \\
	&\left. + (1 - P_{b,1}(h)) \left.\frac{d \Delta G^{ub}_{\text{lig},\text{cnf}}(h)}{d h}\right\vert_{h} \right]
	\label{eqn:ForceReal}
\end{align}
The quantity $P_{b,1}(h)$ is the probability that a single ligand is bound to a receptor when the multivalent particle is at height $h$:
\begin{equation}
	P_{b,1}(h) = \frac{e^{-\Delta G^b_{\text{lig}}(h)/RT }}{e^{-\Delta G^{ub}_{\text{lig}} (h)/RT} + e^{-\Delta G^b_{\text{lig}}(h)/RT }}.
	\label{eqn:PBoundOne}
\end{equation}
In Appendix \ref{app:BellComparison} we demonstrate that this model reproduces the force-dependent bond failure rate anticipated by the Bell model \cite{Bell:1978hj}

Figure \ref{fig:ForceCurves} presents a series of force-extension curves for multivalent binding, predicted by Eq. \ref{eqn:ForceReal}, using various choices of ligand spring constant $k$, rest length $l^\circ$, and effective binding strength $c_R K_{\text{eq}}$. All curves present qualitatively similar behaviour: the restoring force increases roughly linearly with displacement $\Delta z$ from the equilibrium binding height $h_{\text{min}}$. At a critical displacement $\Delta z^* = h^* - h_{\text{min}}$, the force-extension curve reaches a maximum value $F(h^*) = F^*$. 

To understand the physical meaning of $F^*$ we imagine carrying out a force experiment on a single multivalent particle, in which we gradually ramp up the applied force $F_{\text{pull}}$ on the particle. Eventually, the applied force $F_{\text{pull}}$ will exceed the maximum restoring force $F^*$ in the force response function $F(h)$.

At this force, the particle will spontaneously dissociate from the receptor surface. This is analogous to the value of the applied force (stress) at which an elastic material fails in a loading experiment. The quantity $F^*$ shall therefore be referred to as the ``rupture force'' for the multivalent particle, and the displacement height $\Delta z^*$ at which this occurs will be referred to as the ``rupture height''.

The rupture force depends on the design of the multivalent particle. Equation \ref{eqn:ForceReal} can be solved numerically to determine this quantity for any choice of the multivalent design parameters, given a receptor surface density $c_R$. However, in Appendix \ref{app:Approximations} we derive the scaling behaviour of the rupture height and force for ligands that are strong-binding:
\begin{align}
	&F^* \propto N_L \sqrt{2 k RT \ln{\left(\frac{c_R K_{\text{eq}}}{l^\circ}\right)}} \label{eqn:CriticalForceApprox} \\
	&h^* - l^\circ = \Delta z^* \propto \sqrt{\frac{2 RT}{k} \ln{\left(\frac{c_R K_{\text{eq}}}{l^\circ }\right)}} \label{eqn:CriticalDisplacementApprox}
\end{align}
These expressions provide physical insight into how the design of a multivalent particle influences its rupture force.

The numerical calculations in Figure \ref{fig:ForceCurves} reveal the trends predicted by Eqs. \ref{eqn:CriticalForceApprox} and \ref{eqn:CriticalDisplacementApprox}. Stronger-binding ligands or a larger density of surface receptors (i.e. increasing $c_R K_{\text{eq}}$) leads to a larger required pushing force $F^*$ to rupture the particle from the surface. The displacement distance $\Delta z^*$ at which rupture occurs also becomes larger. The left-hand panels of Figure \ref{fig:ForceCurves} reveal that ligands with a shorter rest length $l^\circ$ serve to increase the overall rupture force of the multivalent particle, though this effect is rather small. On the other hand, changing the stiffness $k$ of the ligands has a substantial effect on the rupture force and position, as indicated by the right-hand panels in Figure \ref{fig:ForceCurves}. Less extensible ligands, i.e. those with a larger $k$, lead to a much sharper force-extension curve, a larger required rupture force $F^*$, but a shorter displacement $\Delta z^*$ at which rupture occurs.

\section{Using Force to Obtain Enhanced Super-selective and Hyper-selective Binding} 

\begin{figure*}
	\centering
	\includegraphics[width= 0.90\textwidth]{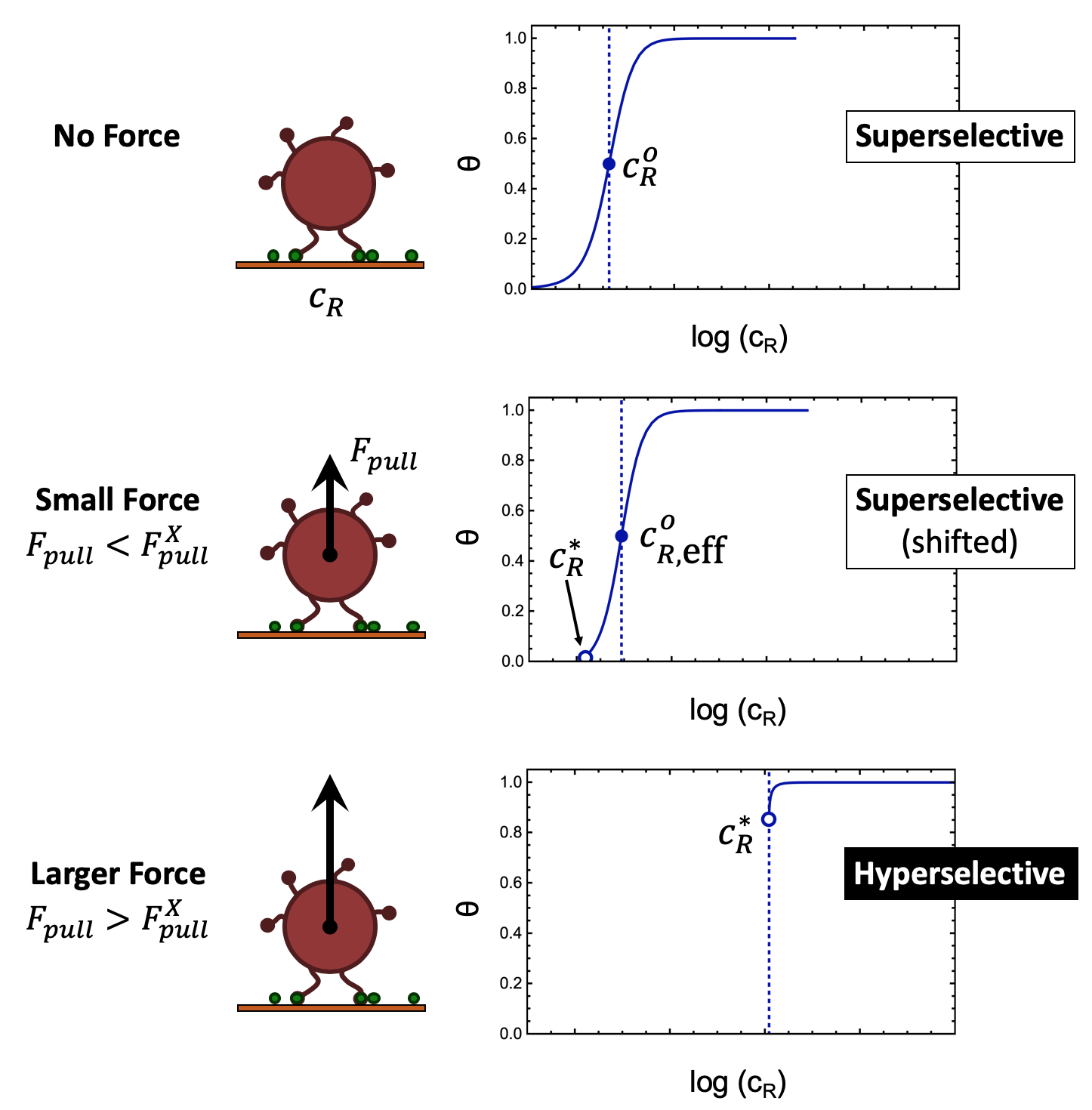}
	\caption{Illustration of the three regimes of multivalent binding discussed in this work, depending on the applied force $F_{\text{pull}}$. Details are discussed in text.}
	\label{fig:BindingRegimes}
\end{figure*} 

Applying a constant force to bound multivalent particles fundamentally alters their surface adsorption/desorption behaviour. Depending on the magnitude of the applied force, the binding transition can be tuned from the standard continuous super-selective profile, to one that is \emph{first-order} and \emph{discontinuous}---a new regime which we term ``hyper-selective''. This is illustrated in Figure \ref{fig:BindingRegimes}. 

The microscopic physics leading to enhanced super-selective, and hyper-selective, binding are now detailed. Example calculations are all performed at the nano-meter length scale, so that the model length scale $b = 1$ nm.

For mathematical simplicity, thermal fluctuations in particle position normal to the substrate are ignored. For example, when the particle is at a height $h$ above the receptor surface, thermal fluctuations will cause the particle to explore an interval of normal positions $\delta h$ around that height coordinate $h$. We neglect these fluctuations, though noting that the fluctuations grow smaller for larger multivalent particles with many simultaneous ligand-receptor bonds. The qualitative influence of these fluctuations on the multivalent adsorption profile are discussed in a subsequent section.

\subsection{Equilibrium super-selective binding transition under no force}

\begin{figure*}
	\centering
	\includegraphics[width= 0.45\textwidth]{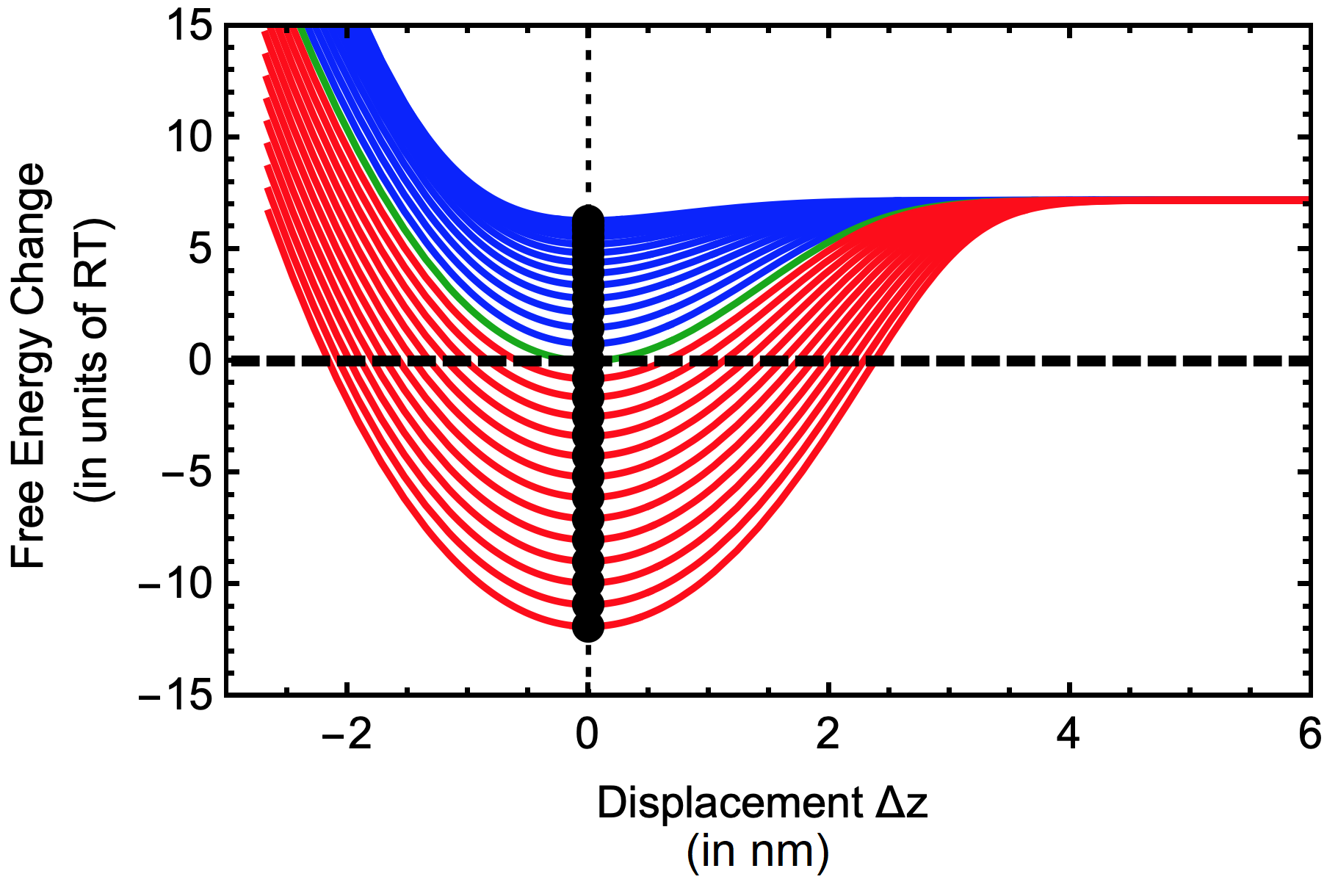}
	\caption{Multivalent binding free energy per particle (Eq. \ref{eqn:FEMainTextWithMu}) as a function of relative distance $\Delta z = h - h_{\text{min}}$ of the particle from the receptor surface, where $h_{\text{min}}$ is the minimum of the curve. Curves are shown for various choices of surface receptor density $c_R$ with zero applied force. Curves coloured blue and red are for choices of receptor density $c_R$ that are respectively before and after the multivalent adsorption transition. The green curve is for the intrinsic transition receptor density $c^\circ_{R,\text{eff}}$. Black dots indicate the equilibrium binding free energy for each $c_R$. Fixed parameters for these calculations are: $l^\circ = 3$ nm, $k = 1$ RT/nm$^2$, $N_L = 5$ ligands, $K_d = 1 / K_{eq} = 100$ $\mu$M, and a molar concentration of $10$ $\mu$M (from which the chemical potential is calculated via Eq. \ref{eqn:Molarity}).}
	\label{fig:FEUnderForceCurvesEQ}
\end{figure*} 

\begin{figure*}
	\centering
	\subfigure[]{\includegraphics[width= 0.45\textwidth]{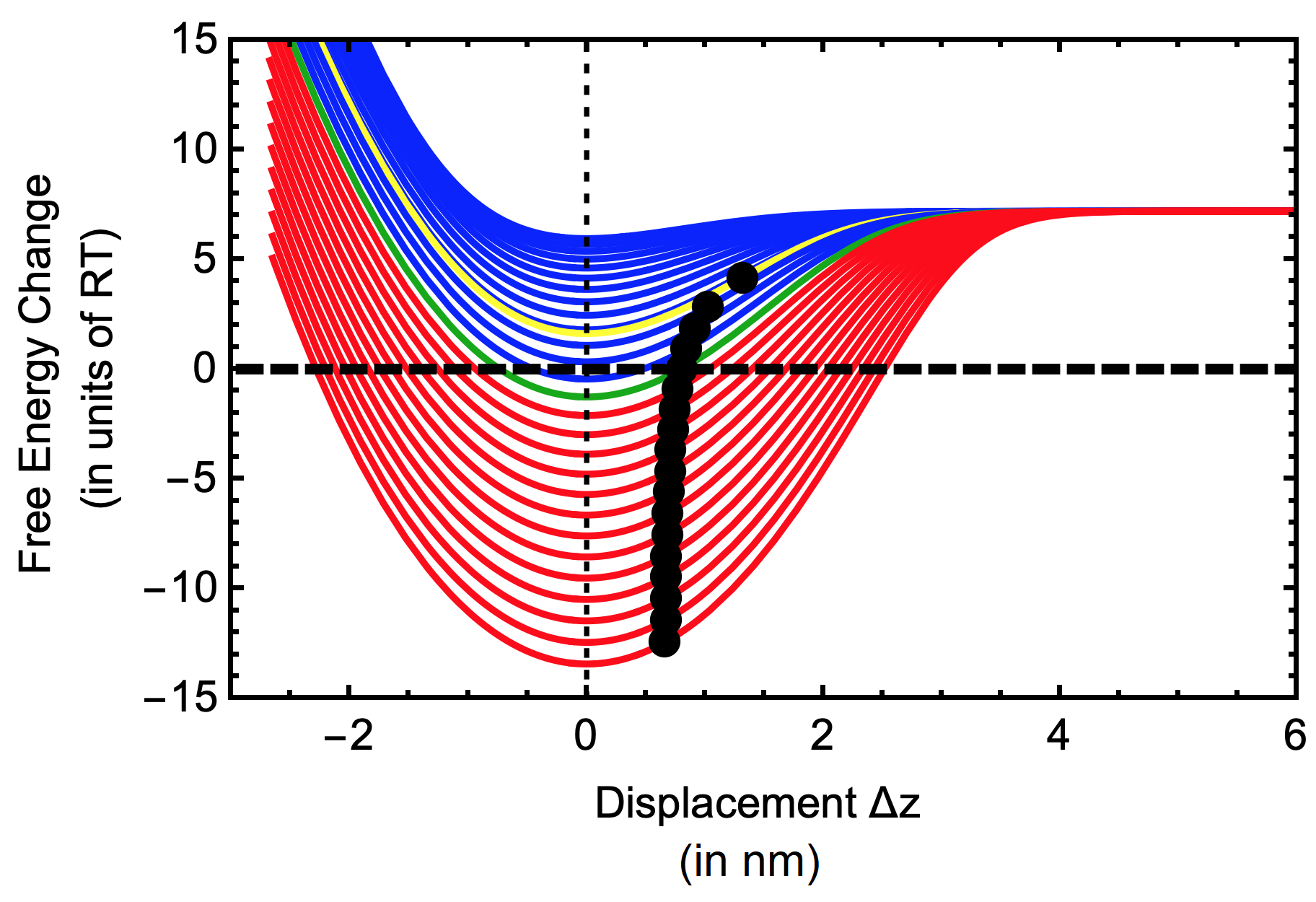}}
	\subfigure[]{\includegraphics[width= 0.45\textwidth]{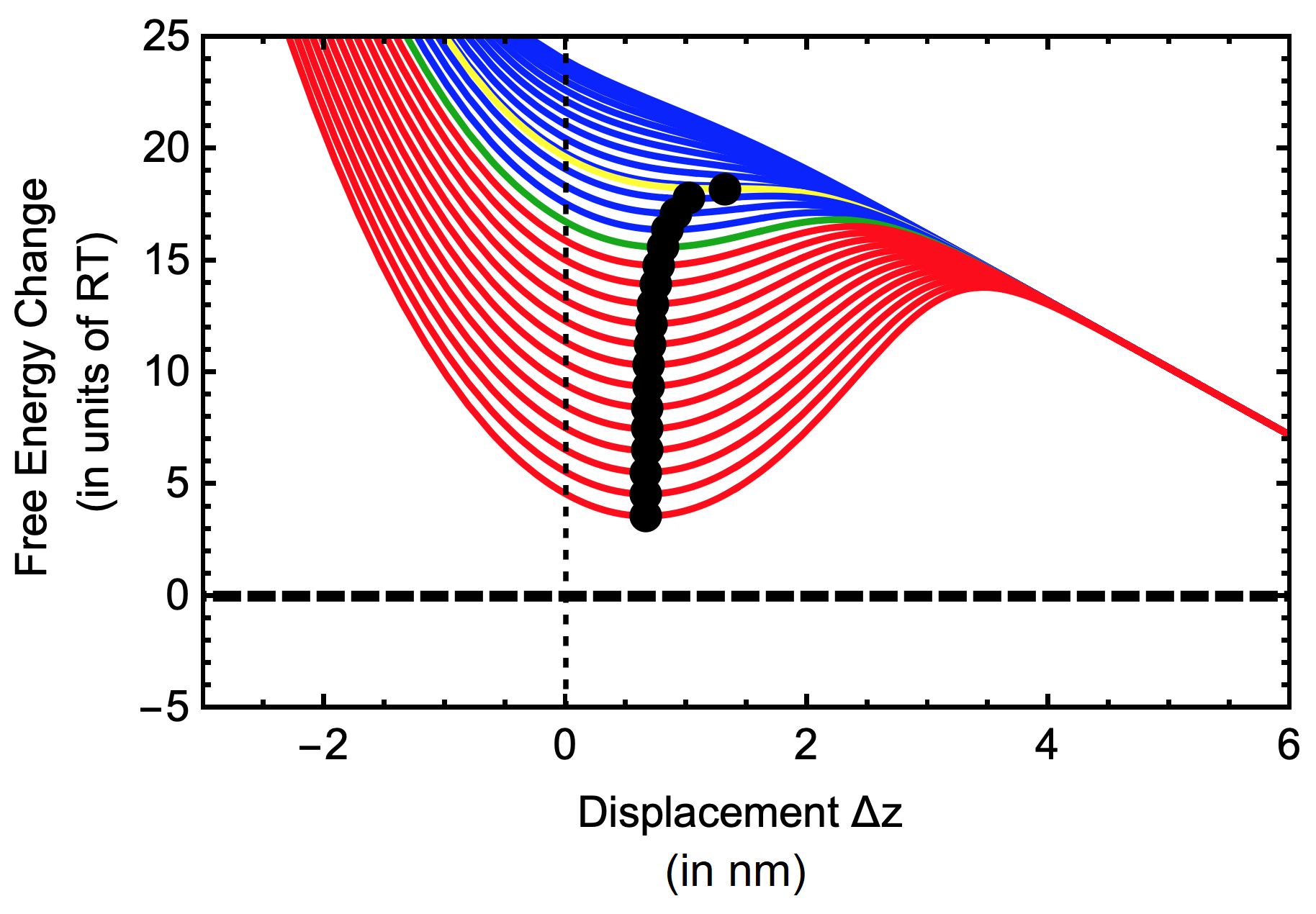}}
	\caption{Multivalent binding free energy per particle (Eq. \ref{eqn:FEMainTextWithMu}) without (a) and with (b) the force field contribution given by Eq. \ref{eqn:ForceFieldFEContribution}, as a function of relative distance $\Delta z = h - h_{\text{min}}$ of the particle from the receptor surface. Curves are shown for various choices of surface receptor density $c_R$, with a low applied force of $F_{\text{pull}} = 3$ $RT/$nm. Curves coloured blue, red, and green are defined in the caption to Figure \ref{fig:FEUnderForceCurvesEQ}. The yellow curve is for the mechanical transition receptor density $c_R^*$. Black dots indicate the equilibrium binding free energy for each $c_R$. Curves without a black dot correspond to choices of $c_R < c_R^*$. See Figure \ref{fig:FEUnderForceCurvesEQ} caption for fixed parameters in these calculations.}
	\label{fig:FEUnderForceCurvesForce}
\end{figure*} 

\begin{figure*}
	\centering
	\subfigure[]{\includegraphics[width= 0.45\textwidth]{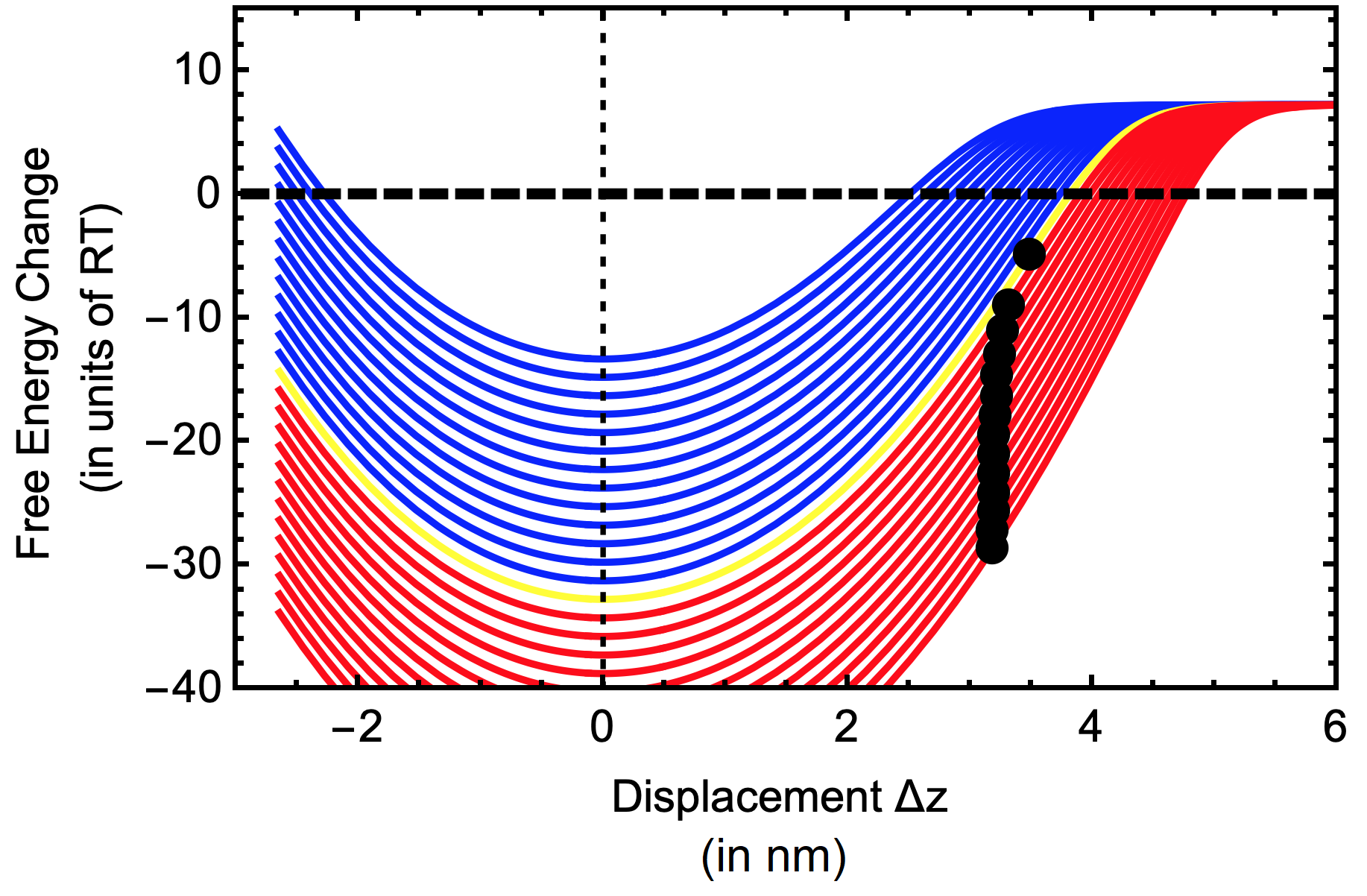}}
	\subfigure[]{\includegraphics[width= 0.45\textwidth]{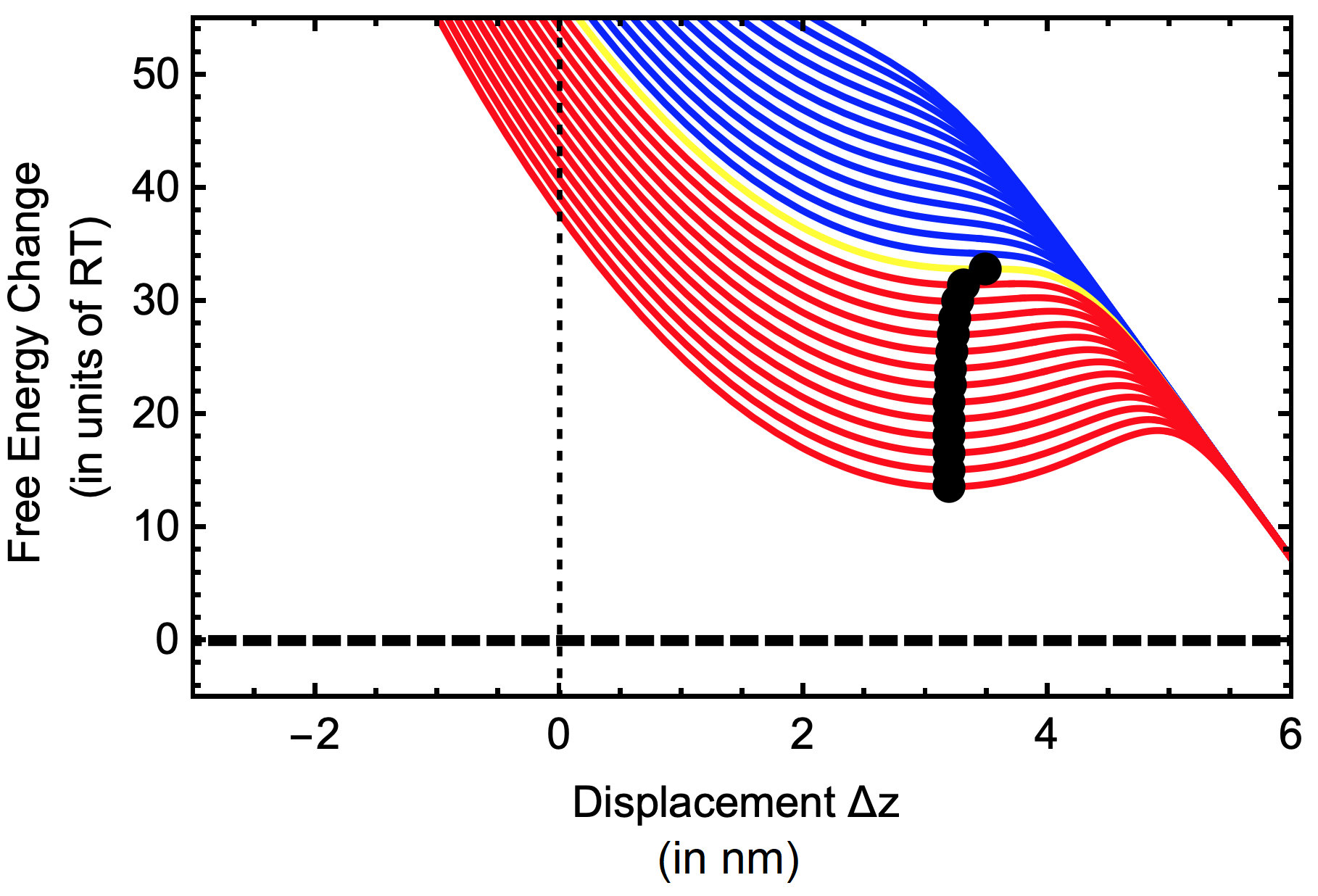}}
	\caption{Multivalent binding free energy per particle (Eq. \ref{eqn:FEMainTextWithMu}) without (a) and with (b) the force field contribution given by Eq. \ref{eqn:ForceFieldFEContribution}, as a function of relative distance $\Delta z = h - h_{\text{min}}$ of the particle from the receptor surface. Curves are shown for various choices of surface receptor density $c_R$, with a high applied force of $F_{\text{pull}} = 15$ $RT/$nm. Curves coloured blue, red, and green are defined in the caption to Figure \ref{fig:FEUnderForceCurvesEQ}. The yellow curve is for the mechanical transition receptor density $c_R^*$. Black dots indicate the equilibrium binding free energy for each $c_R$. Curves without a black dot correspond to choices of $c_R < c_R^*$. See Figure \ref{fig:FEUnderForceCurvesEQ} caption for fixed parameters in these calculations.}
	\label{fig:FEUnderForceCurvesHigherForce}
\end{figure*} 

Consider a solution of multivalent particles, with a given concentration $[M]$ (in mol/$b^3$), in contact with a substrate with receptors at a surface molar density of $c_R$ (in mol/$b^2$). The multivalent particles have a diameter of $a$, such that their excluded volume is $V_{ex} = a^3$ (in units of $b^3$), and the amount of area they occupy when bound to the substrate is $A_{ex} = a^2$. For all examples here we choose the particle diameter to be $a = 5 \text{ nm}$.

Let the chemical potential for the particles in solution, corresponding to the molarity $[M]$, be $\mu$. If the molar concentration $[M]$ is dilute, then the chemical potential for the particles in solution is approximately
\begin{equation}
	\frac{\mu}{RT} \approx \ln{([M] N_{\rm A} V_{ex})},
	\label{eqn:Molarity}
\end{equation}
where $N_{\rm A}$ is Avogadro's number. Here we chose the ``natural'' reference concentration for this system $c_{0}=1/N_{\rm A}V_{ex}$, while usually the standard reference is taken: $c_{\rm 0, std}=1\,$M. A simple rescaling operation: $\mu = \mu_{\rm std} + \ln(c_{\rm 0,std} V_{ex} N_{\rm A})$ connects the two definitions. 

For purposes of clarity, we also introduce the ``surface receptor count''
\begin{equation}
	N_R \equiv c_R A_{ex} N_{\rm A}
	\label{eqn:NREquation}
\end{equation}
as the average number of receptors that a bound multivalent particle can simultaneously reach. This is the measure of receptor density we employ for the figures in this section.

The chemical potential $\mu$ shifts the binding free energy of the multivalent particles, Eq. \ref{eqn:FEMainText}, by an additive constant, leading to the net binding free energy
\begin{align}
	\Delta G(h) = \Delta G_{\text{bind}}(h) - \mu.
	\label{eqn:FEMainTextWithMu}
\end{align}
Figure \ref{fig:FE} presented examples of these curves, revealing that they have a distinct minimum $\Delta G(h_{\text{min}})$ at the equilibrium binding position $h_{\text{min}}$. This value will be referred to as $\Delta G^{\text{min}}$. Changing the chemical potential $\mu$, all other parameters being fixed, adjusts the depth and sign of the minimum $\Delta G^{\text{min}}$. For large negative $\Delta G^{\text{min}}$, the multivalent particles bind strongly and spontaneously, while for positive $\Delta G^{\text{min}}$ binding vanishes. Indeed, the fraction $\theta$ of the receptor surface occupied by bound particles is determined by the standard Langmuir isotherm:
\begin{align}
	\theta &= \frac{e^{-\Delta G^{\text{min}}/RT}}{1 + e^{-\Delta G^{\text{min}}/RT}}.
	\label{eqn:ThetaEq}
\end{align}
For large positive $\Delta G^{\text{min}}$, $\theta \rightarrow 0$, while in the opposite limit $\theta \rightarrow 1$. Since the equilibrium binding free energy $\Delta G^{\text{min}}$ changes continuously with $\mu$, then the adsorption transition is \emph{continuous}. The binding curve $\theta$ as a function of $\mu$ has the characteristic continuous sigmoidal shape, and its inflection point occurs near the choice of $\mu$ where $\Delta G^{\text{min}} = 0$.

In the complementary sense, $\mu$ can be fixed and the surface receptor density $c_R$ can be varied. Given some choice of design for the multivalent particles, the chemical potential $\mu$ defines the critical receptor density $c_R^\circ$ where the inflection point of $\theta(c_R)$ occurs. 

\begin{figure*}
	\centering
	\includegraphics[width= 0.70\textwidth]{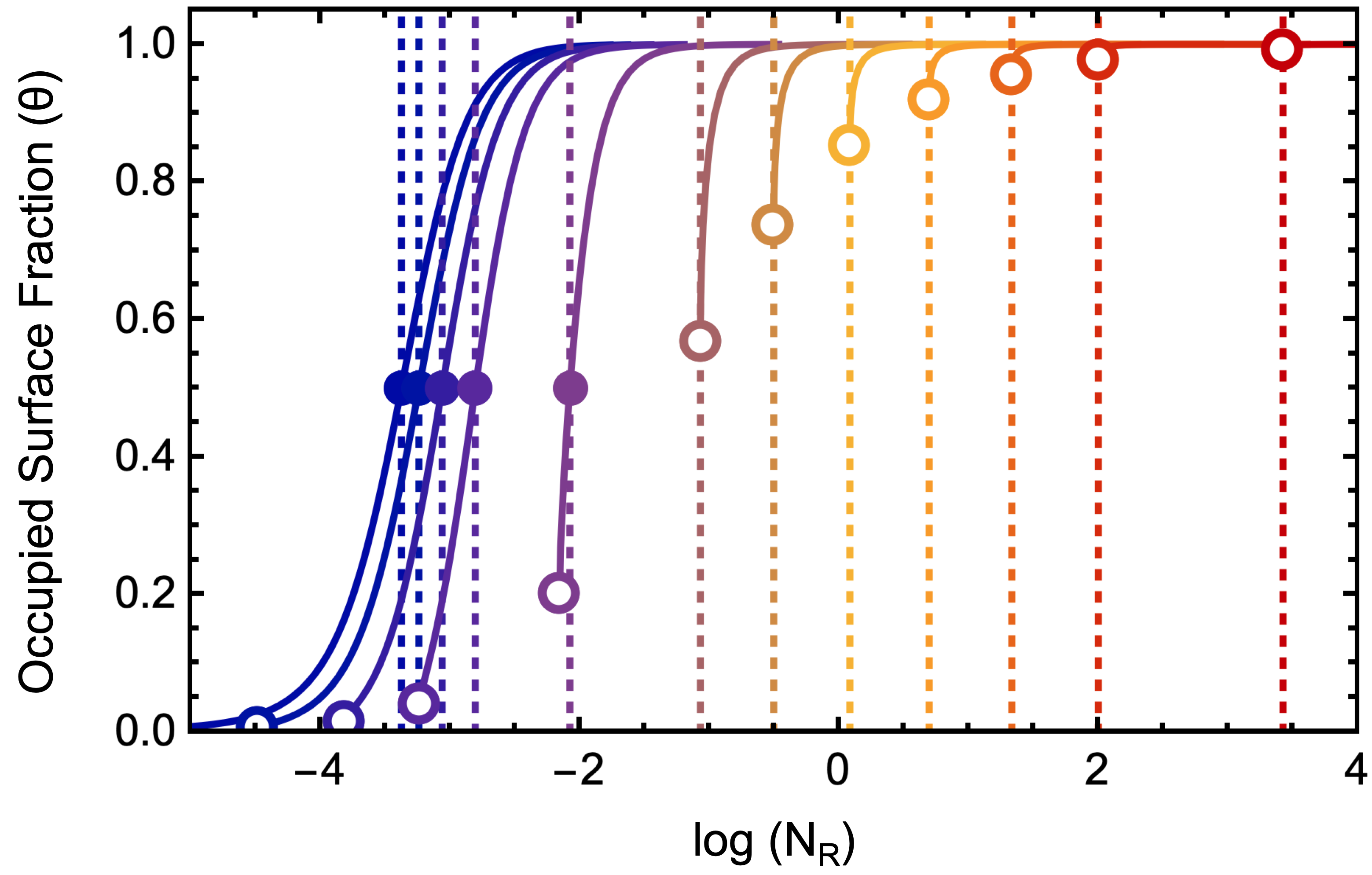}
	\caption{Multivalent particle adsorption isotherms $\theta$, given by Eqs. \ref{eqn:ThetaEq} and \ref{eqn:ThetaShift}, as a function of the natural logarithm of the surface receptor count $N_R$ (related to $c_R$ via Eq. \ref{eqn:NREquation}). Results are given for various choices of $F_{\text{pull}} = 0$ $RT/$nm (blue curve) to $15$ $RT/$nm (red curves). Vertical dashed lines indicate the threshold values of $c_R$ for multivalent adsorption at each given applied force. For choices of $F_{\text{pull}} > 0$ where $c_R^* < c^\circ_{R,\text{eff}}$, the adsorption profile displays two points of interest. The solid point indicates the intrinsic threshold receptor density $c^\circ_{R,\text{eff}}$, corresponding to where $\theta = 1/2$. The open point indicates the mechanical threshold receptor density $c_R^*$---below which \emph{no} particles bind to the surface---and the value of $\theta$ at that receptor density. For larger pulling forces, the intrinsic threshold receptor density $c^\circ_{R,\text{eff}}$ (solid point) vanishes, and the multivalent binding transition happens at $c_R^*$ (open point) in a discontinuous first-order fashion. Fixed parameters for these calculations are: $l^\circ = 3$ nm, $k = 1$ RT/nm$^2$, $N_L = 5$ ligands, $K_d = 1 / K_{eq} = 100$ $\mu$M, and a molar concentration of $[M] = 10$ $\mu$M.}
	\label{fig:ThetaVsCR}
\end{figure*} 

For example, Figure \ref{fig:FEUnderForceCurvesEQ} displays a series of binding free energy curves $\Delta G(h; c_R)$ for different $c_R$, all for the same multivalent particle design and chemical potential $\mu$. In each curve, the minimum is indicated by a black dot. Curves where the minimum is greater than zero are shown in blue, while those that are less than zero are red. The green curve is for the choice $c_R = c_R^\circ$, where the minimum binding free energy is exactly equal to zero, corresponding to $\theta(c_R^\circ) = 1/2$. The binding profile $\theta(c_R)$ is shown in full as the left-most blue curve in Figure \ref{fig:ThetaVsCR}. Since the minimum binding free energy passes continuously through zero as a function of $c_R$, then the adsorption transition shown in Figure \ref{fig:ThetaVsCR} is continuous. 

\subsection{Shifted super-selective binding transition under weak force}

Applying a constant force $F_{\text{pull}}$ to multivalent particles at equilibrium leads to \emph{two} new kinds of control over the adsorbed amount $\theta$:
\begin{enumerate}
	\item{$F_{\text{pull}}$ shifts the equilibrium binding free energy of the multivalent particles to higher (less negative) values, so that $\theta$ is lower for a given receptor density $c_R$;}
	\item{Multivalent particles are only able to bind when the surface receptor density is sufficiently large, such that the applied force $F_{\text{pull}}$ is \emph{smaller than} the rupture force $F^*(c_R)$.}
\end{enumerate}
Let us initially ignore the latter condition, which can be safely done when $F_{\text{pull}}$ is small. 

As noted in the Introduction, the force field can be applied: only to multivalent particles when they are bound; or to all particles regardless of whether they are bound or unbound. These are two distinct regimes which lead to markedly different adsorption/desorption behaviour. 

To begin, the case where only bound multivalent particles are exposed to the force field is examined. Applied force causes bound multivalent particles to move upward in their free energy landscape, to a new equilibrium coordinate $\tilde{h}_{eq} \leq h^*$ where the \emph{gradient} of the binding free energy $d \Delta G(h; c_R) / dh$ is equal to $F_{\text{pull}}$. The value of the free energy at $\tilde{h}_{eq}$, given as $\Delta G (\tilde{h}_{eq}; c_R) \equiv \Delta G^{\text{eff}}(c_R, F_{\text{pull}})$ is now the equilibrium binding free energy of the multivalent particles within the force field. 

Figure \ref{fig:FEUnderForceCurvesForce}a presents free energy curves for the same design parameters as in Figure \ref{fig:FEUnderForceCurvesEQ}, illustrating how an applied force pushes the equilibrium binding free energy away from $\Delta G^{\text{min}}(c_R)$, to the new value $\Delta G^{\text{eff}}(c_R, F_{\text{pull}})$. These equilibria are indicated by the black dots in the figure for each receptor density $c_R$. The multivalent adsorption transition now occurs at the choice of $c_R$ where the \emph{effective} binding free energy $\Delta G^{\text{eff}}(c_R, F_{\text{pull}}) = 0$, plotted as the green curve in Figure \ref{fig:FEUnderForceCurvesForce}a. This new critical receptor density is denoted ``$c^\circ_{R,\text{eff}}$'', and it is larger than $c_R^\circ$.

The fractional coverage of particles on the surface is calculated by
\begin{align}
	\theta &= \frac{e^{-\Delta G^{\text{eff}}(c_R, F_{\text{pull}})/RT}}{1 + e^{-\Delta G^{\text{eff}}(c_R, F_{\text{pull}})/RT}}.
	\label{eqn:ThetaShift}
\end{align}
The adsorption profile $\theta(c_R)$ for the parameters employed in Figure \ref{fig:FEUnderForceCurvesForce}a is shown as the third blue curve from the left in Figure \ref{fig:ThetaVsCR}. The applied force has shifted the adsorption inflection point to the larger receptor density $c^\circ_{R,\text{eff}}$, though it largely resembles the adsorption transition at zero force. The only notable difference is the appearance of a second point of interest indicated by the open circle, the subject of the next section.

Since only bound multivalent particles are exposed to the force field in this regime, then the unbound particles in the reservoir need not do any work against the force field in order to approach and bind to the surface. The applied force therefore does not influence the shape of the free energy profiles $\Delta G (\tilde{h}_{eq}; c_R)$ in Figure \ref{fig:FEUnderForceCurvesForce}a. The force only serves to change the equilibrium binding height, and the equilibrium binding free energy from $\Delta G^{\text{min}}(c_R)$ to $\Delta G^{\text{eff}}(c_R, F_{\text{pull}})$.

Let us now examine the scenario where both bound and unbound multivalent particles are exposed to the force field. The simulations presented in the Introduction in Figure \ref{fig:TineSims} provide a practical example of this. Repulsive polymers were tethered onto the multivalent particle cores, leading to a repulsive normal force when the particles approach or are bound to the receptor surface.

In this case, the binding free energy for a multivalent particle at distance $h$ from the surface is shifted by a contribution from the constant force field of the form
\begin{equation}
	\frac{\Delta G_{field}}{RT} = -F_{\text{pull}} (h - h_{bulk}).
	\label{eqn:ForceFieldFEContribution}
\end{equation}
Here, $h_{bulk}$ is a reference height which defines the bulk chemical potential. For example, we can imagine $h_{bulk}$ to be the height above the surface where the force field begins. For all $h < h_{bulk}$, both bound and unbound particles feel $F_{\text{pull}}$, while for $h > h_{bulk}$ the force field is zero and we recover the bulk chemical potential $\mu$.

Including this field contribution into the overall multivalent binding free energy yields the curves shown in Figure \ref{fig:FEUnderForceCurvesForce}b. In those results we have defined ``bulk'' to be at $h_{bulk} - h_{min} = 6$, the right-most horizontal coordinate in the figure. We clearly see that the bound-state free energy equilibria $\Delta G^{\text{eff}}(c_R, F_{\text{pull}})$ for the multivalent particles within the force field correspond to the local minima in the curves in Figure \ref{fig:FEUnderForceCurvesForce}b.

According to Figure \ref{fig:FEUnderForceCurvesForce}b, exposing the unbound particles to the force field effectively makes their bound-state free energies more positive (i.e. less favourable) relative to the chemical potential of the particles in bulk. This is because an unbound particle must first ``pay'' the thermodynamic free energy cost for approaching the surface within the force field before it is able to form ligand-receptor bonds. (On the other hand, if the force field only applies to particles that are bound by one or more ligands at the surface, then the bound particles retain their intrinsic free energy landscapes shown in Figure \ref{fig:FEUnderForceCurvesForce}a.)

The result of including the force field on the unbound particles, according to the free energy calculations in Figure \ref{fig:FEUnderForceCurvesForce}b, is that the adsorption profile shifts to even larger values of $c_R$. In fact including the force field for the unbound particles actually delays the adsorption transition to values of $c_R$ well greater than those included in the figure. None of the values of $c_R$ in Figure \ref{fig:FEUnderForceCurvesForce}b yield an equilibrium binding free energy at the local minimum that is less than zero. We also saw the shift in the adsorption transition to larger $c_R$ in the simulation results in Figure \ref{fig:TineSims}, consistent with the theoretical analysis here.

\subsection{Force-induced mechanical transition point \& hyper-selective binding}

An external force imposes the strict condition that particles only bind (or remain bound) to the receptor surface if the rupture force $F^*(c_R)$ for the given receptor density $c_R$ is \emph{greater than} the external force $F_{\text{pull}}$. This condition results in an additional critical value of receptor concentration, which we will call the ``mechanical'' transition point $c_R^*$. Below $c_R^*$, $F^*(c_R) < F_{\text{pull}}$ so that no multivalent binding is allowed \emph{regardless of the magnitude of the binding free energy $\Delta G^{\text{eff}}(c_R, F_{\text{pull}})$}. Above $c_R^*$, we have $F^*(c_R) > F_{\text{pull}}$ so that multivalent binding is permitted.

\begin{figure}
	\centering
	\includegraphics[width= 0.49\textwidth]{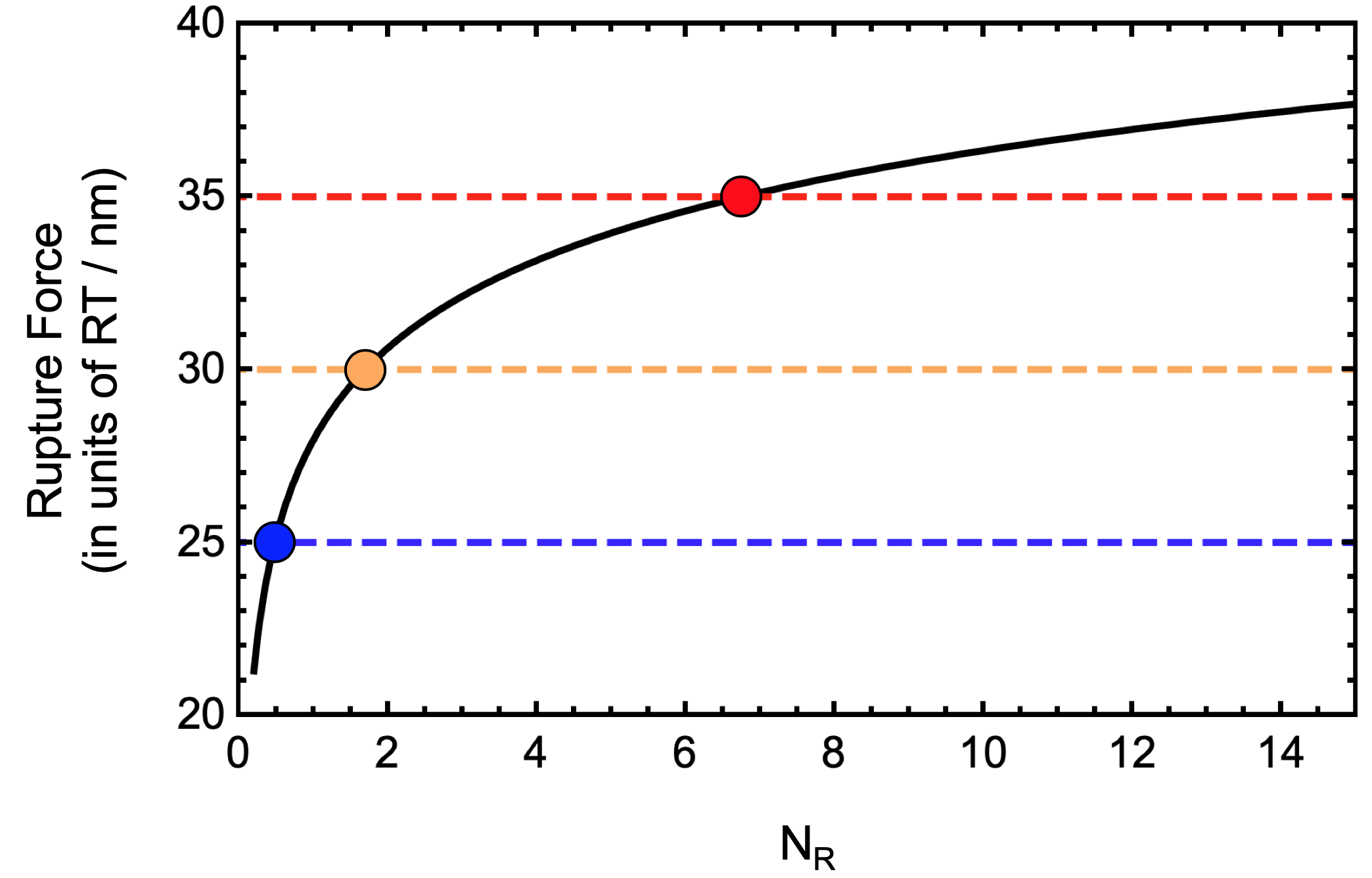}
	\caption{Rupture force $F^*$ as a function of surface receptor count $N_R$ (solid black curve). Three arbitrary examples of applied forces $F_{\text{pull}}$ are shown by the coloured dashed lines. The coordinate $c_R$ where a horizontal line intersects with the $F^*(c_R)$ curve defines the mechanical transition receptor density $c_R^*$ for that given applied force $F_{\text{pull}}$. Rupture force curve is computed for multivalent particles with: $l^\circ = 3$ nm, $k = 6$ RT/nm$^2$, $N_L = 5$ ligands, and $K_{d} = 1 / K_{eq} = 100$ $\mu$M.}
	\label{fig:FStarVsCR}
\end{figure} 

Figure \ref{fig:FStarVsCR} presents this idea graphically. The plot shows, for a given choice of multivalent design parameters $(l^\circ, K_{eq}, k, N_L)$, how the rupture force $F^*(c_R)$ for a bound multivalent particle varies with the density $c_R$ of receptors on the surface. This generally follows Eq. \ref{eqn:CriticalForceApprox} as derived in Appendix \ref{app:Approximations}, i.e. the rupture force varies with the square-root of the logarithm of $c_R$. The plot also contains three examples of possible applied forces $F_{\text{pull}}$, given as horizontal dashed lines. The intersection coordinate between these lines and the $F^*(c_R)$ define the mechanical transition receptor density $c_R^*$ for the given $F_{\text{pull}}$.

Thus, for a given choice of nonzero $F_{\text{pull}}$, there is the critical receptor density $c_R^*$ below which no multivalent binding can occur. 

Let us examine this mechanical stability threshold for systems where the applied force field only affects bound particles. For the force field magnitude examined in the previous section, the free energy curve for $c_R = c_R^*$ is plotted in yellow in Figure \ref{fig:FEUnderForceCurvesForce}a. The free energy curves for smaller values of $c_R$ lack an equilibrium point marker (black dot), i.e. nowhere along those curves is the derivative $d \Delta G (h; c_R)/dh = 0$, and so multivalent binding does not occur.

The effect of this is to ``truncate'' the multivalent adsorption profiles in Figure \ref{fig:ThetaVsCR} when force is applied. The open circles indicate the coordinate $c_R^*$ (and corresponding value of $\theta$) below which binding is prohibited. For low applied force (blue curves), this has only a minor influence on the adsorption profile; truncation only occurs well below the intrinsic inflection point $c^\circ_{R,\text{eff}}$.

On the other hand, applying an increasingly larger $F_{\text{pull}}$ causes the truncation point $c_R^*$ to creep up the adsorption profile in Figure \ref{fig:ThetaVsCR}. In doing so, $c_R^*$ defines a \emph{discontinuous} jump in the adsorbed amount $\theta$, from $0$ to a non-zero value.

For sufficiently large force, this truncation point progresses further along and entirely overtakes the intrinsic transition $c^\circ_{R,\text{eff}}$. This we refer to as the ``crossover'' point, where the intrinsic binding threshold $c^\circ_{R,\text{eff}}$ ceases to exist, in lieu of the mechanical binding transition point $c_R^*$. Let $F_{\text{pull}}^{X}$ be the unique value of the pulling force where this crossover occurs, and $c_{R}^X \equiv c_R^* = c^\circ_{R,\text{eff}}$ be the value of the adsorption threshold receptor density at this crossover.

Here, the adsorption of the multivalent particles becomes very much like a step-function in receptor density space, with a critical point at $c_R^*$. This is clear in the yellow and red adsorption profiles in Figure \ref{fig:ThetaVsCR}. For values of $c_R$ only infinitesimally below $c_R^*$, the free energy per multivalent particle is the bulk value, $\mu$, and thus the adsorbed amount $\theta = 0$. Subsequently, right at $c_R*$, there is a sudden jump in the free energy per particle to $\Delta G^{\text{eff}}(c_R^*, F_{\text{pull}})$, corresponding to the fact that $F^*(c_R)$ is now greater than $F_{\text{pull}}$. This causes an instantaneous jump in the adsorbed amount $\theta$ to the value
\begin{equation}
	\theta(c_R^*) = \frac{e^{-\Delta G^{\text{eff}}(c_R^*, F_{\text{pull}})/RT}}{1 + e^{-\Delta G^{\text{eff}}(c_R^*, F_{\text{pull}})/RT}}.
\end{equation}
We refer to this discontinuous transition as ``hyper-selective'' multivalent binding, in order to distinguish it from the standard continuous super-selective transitions under weak or no applied force.

To better understand this feature, Figure \ref{fig:FEUnderForceCurvesHigherForce}a presents free energy profiles for a choice of $F_{\text{pull}}$ where binding is in the hyper-selective regime, keeping all other parameters the same as in Figures \ref{fig:FEUnderForceCurvesEQ} and \ref{fig:FEUnderForceCurvesForce}a. The green ($c^\circ_{R,\text{eff}}$) transition point has vanished, and now only the yellow $c_R^*$ transition remains. This transition point occurs rather deep in the free energy landscape. Upon reaching $c_R^*$, the position of the binding equilibrium $\tilde{h}_{eq}^*$ is such that the free energy $\Delta G^{\text{eff}}(c_R^*, F_{\text{pull}})$ is already substantially non-zero and negative. As a result, for this choice of $F_{\text{pull}}$ the mechanical transition $c_R^*$ defines the binding transition of the multivalent particles.

As noted earlier, thermal fluctuations in the multivalent particles' vertical coordinates $h$ are not incorporated mathematically in this discussion. However, we can make a semi-quantitative assessment of how they will influence the first-order binding regime. When the multivalent particles are small, and the total binding free energy is on the order of $RT$, then thermal fluctuations will tend to blur the transition back into a second-order process. On the other hand, for larger particles with larger binding free energies, the influence of these thermal fluctuations will diminish.

When the force field is applied to both bound and unbound particles, the hyper-selective transition is lost. This can be seen by examining the free energy profiles shown in Figure \ref{fig:FEUnderForceCurvesHigherForce}b, now including the force field contribution. Like in Figure \ref{fig:FEUnderForceCurvesForce}b, including the force field on the unbound particles shifts their bound-state free energies to more positive values, leading to a shift in the adsorption transition to larger $c_R$. The hyper-selective binding transition is lost, since now the equilibrium binding free energy for $c_R^*$ is well above zero.

This underscores a more general point: the hyper-selective binding regime is only obtained under conditions where the equilibrium binding free energy (relative to the bulk chemical potential) at $c_R^*$ is \emph{less than} zero. This is the case in Figure \ref{fig:FEUnderForceCurvesHigherForce}a, when the force field is only applied to bound particles. However, when the force field is applied to both bound and unbound particles as in Figure \ref{fig:FEUnderForceCurvesHigherForce}b, the initial thermodynamic work that unbound particles must do against the force field in order to bind to the surface undermines their subsequent adhesion strength to the surface. 

In general, when the force field applies to both bound and unbound particles, there is no choice of parameters where the equilibrium binding free energy is less than zero at $c_R^*$. Therefore, we believe that a hyper-selective transition can only be obtained under non-equilibrium conditions when the force field applies to bound particles only. 

\section{Tuning the Binding Transitions with Particle Design}

The previous section revealed three key transition points for multivalent binding in $c_R$-space, when just the bound particles are placed in a constant force field $F_{\text{pull}}$:
\begin{enumerate}
	\item{Intrinsic transition point $c_R^\circ$: the critical surface receptor density where $\theta = 1/2$ (i.e. where $\Delta G^{\text{min}}(c_R) = 0$) for particles under no external force.}
	\item{Shifted intrinsic transition point $c^\circ_{R,\text{eff}}$: the receptor density where $\theta = 1/2$ (i.e. where $\Delta G^{\text{eff}}(c_R) = 0$) for particles under force. This transition point only exists when $F_{\text{pull}} < F_{\text{pull}}^X$.}
	\item{Mechanical transition point $c_R^*$: the surface receptor density where $F^*(c_R) = F_{\text{pull}}$ for particles under nonzero force. This transition point exists for all $F_{\text{pull}} > 0$.}
\end{enumerate}
This section examines how the transition points vary with the design of the multivalent particles, as well as their concentration $[M]$ in solution.

\begin{figure*}
	\centering
	\includegraphics[width= 1.00\textwidth]{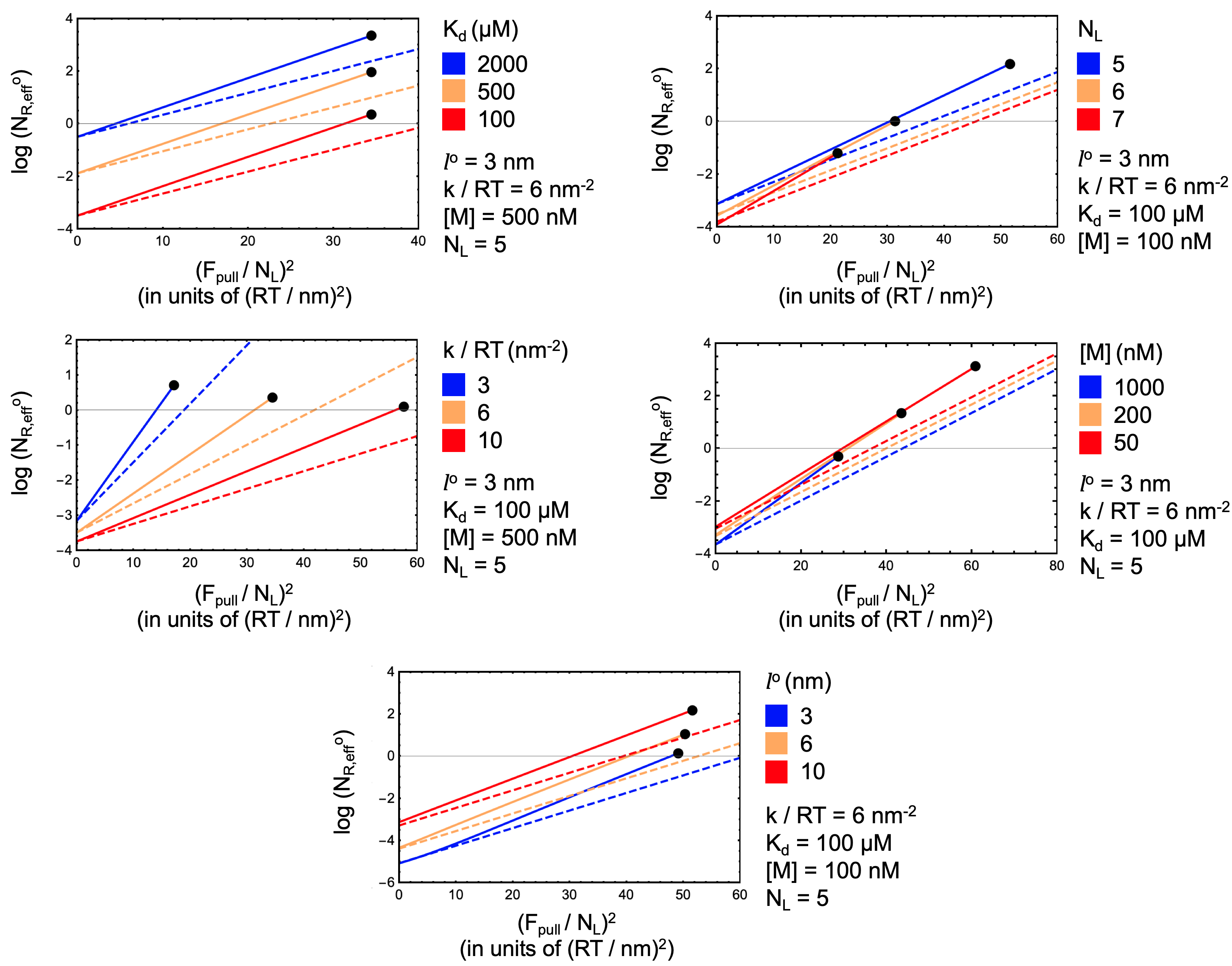}
	\caption{Plots of the logarithm of the intrinsic transition point $c^\circ_{R,\text{eff}}$ (expressed as $N^\circ_{R,\text{eff}}$ via Eq. \ref{eqn:NREquation}) as a function of $(F_{\text{pull}} / N_L)^2$, for various choices of: ligand/receptor dissociation constant $K_d = 1 / K_{eq}$, ligand spring constant $k$, equilibrium ligand length $l^\circ$, number of ligands $N_L$, and multivalent particle molar concentration $[M]$ (related to the chemical potential via Eq. \ref{eqn:Molarity}). Solid lines are numerical calculations, and dashed lines are the predicted scaling according to Eq. \ref{eqn:CRScalingForceIntr}. Black points correspond to the choice of $(F_{\text{pull}}/N_L)^2$ where $c_R^*$ is first equal to $c^\circ_{R,\text{eff}}$: for larger $(F_{\text{pull}}/N_L)^2$, the transition point $c^\circ_{R,\text{eff}}$ no longer exists.}
	\label{fig:CRIntrScalings}
\end{figure*} 

An estimate for how the intrinsic transition point $c_R^\circ$ scales with the chemical potential and multivalent design parameters can be made in the strong-binding ligand limit using Eq. \ref{eqn:FEMainTextApprox}. At zero force, the equilibrium binding height will be near $h_{\text{min}} \approx l^\circ$. Including the chemical potential term introduced in Eq. \ref{eqn:FEMainTextWithMu}, and then invoking the scaling of the chemical potential with the concentration $[M]$ given by Eq. \ref{eqn:Molarity}, leads to
\begin{equation}
	\frac{\Delta G^{\text{min}}(c_R)}{RT} \approx - N_L \ln{\left(\frac{c_R K_{\text{eq}}}{l^\circ}\right)} + \ln{\left(\frac{1}{[M] N_{\rm A} V_{ex}}\right)}
\end{equation}
Solving for the value of $c_R$ where $\Delta G^{\text{min}}(c_R) = 0$ yields
\begin{align}
	&\ln{c_R^\circ} \propto \ln{\left(\frac{l^\circ}{K_{\text{eq}}}\right)} + \ln{\left(\frac{1}{[M] N_{\rm A} V_{ex}}\right)}.
	\label{eqn:CRScalingIntrinsic}
\end{align}
Next, Eq. \ref{eqn:FEMainTextApprox} can be invoked to determine how the binding free energy equilibria scale with the applied force $F_{\text{pull}}$. The equilibrium binding height of the multivalent particles shifts to
\begin{equation}
	\tilde{h}_{eq} \propto \frac{F_{\text{pull}}}{N_L k} + l^\circ,
	\label{eqn:HEqApprox}
\end{equation} 
obtained by combining Eqs. \ref{eqn:CriticalForceApprox} and \ref{eqn:CriticalDisplacementApprox}. Putting this expression in for $h$ in Eq. \ref{eqn:FEMainTextApprox} and again including the chemical potential term yields
\begin{align}
	\frac{\Delta G^{\text{eff}}(c_R, F_{\text{pull}})}{RT} \approx &\frac{F_{\text{pull}}^2}{2 R T N_L k} - N_L \ln{\left(\frac{c_R K_{\text{eq}}}{l^\circ}\right)} \nonumber \\
	& + \ln{\left(\frac{1}{[M] N_{\rm A} V_{ex}}\right)} \label{eqn:FEForGivenPush}.
\end{align}
As before, the binding inflection point in $c_R$-space is the choice of $c_R$ where $\Delta G^{\text{eff}}(c_R, F_{\text{pull}}) = 0$. Thus, we see that $F_{\text{pull}}$ effectively shifts the transition point to a larger value $c^\circ_{R,\text{eff}} > c_R^\circ$:
\begin{equation}
	\ln{c^\circ_{R,\text{eff}}} \propto \ln{c_R^\circ} + \frac{1}{2 k R T} \left(\frac{F_{\text{pull}}}{N_L}\right)^2
	\label{eqn:CRScalingForceIntr}
\end{equation}
In the absence of any force, the second term goes to zero, and we recover the standard scaling of $c_R^\circ$.

Figure \ref{fig:CRIntrScalings} presents a series of plots showing how $c^\circ_{R,\text{eff}}$ varies with the squared force per ligand, $(F_{\text{pull}}/N_L)^2$, for different choices of multivalent particle concentration and design parameters. The choices of $(F_{\text{pull}}/N_L)^2$ where $c^\circ_{R,\text{eff}}$ vanishes (i.e. due to being eclipsed by $c_R^*$) are shown as black dots. These are the ``crossover'' points between the super-selective and hyper-selective binding regimes introduced in the previous section, occurring at the choice of force $F_{\text{pull}}^X$ and located at receptor density $c_R^X$.

Changing $K_{eq}$ or $l^\circ$ shifts $\ln{c^\circ_{R,\text{eff}}}$ by a constant, as predicted by Eq. \ref{eqn:CRScalingForceIntr}. It also leads only to a change in $c_{R}^X$, with very little change in the crossover force $F_{\text{pull}}^{X}$. In contrast, the ligand elasticity $k$ affects the \emph{slope} of $\ln{c^\circ_{R,\text{eff}}}$ with $(F_{\text{pull}}/N_L)^2$. As a result, there is a significant variation in the crossover force $F_{\text{pull}}^{X}$, with only marginal change in the corresponding crossover receptor density $c_{R}^X$. Finally, changing the number of ligands $N_L$ on the particles, or the particle concentration, leads to variation in \emph{both} $F_{\text{pull}}^{X}$ and $c_{R}^X$, but little change in how $\ln{c^\circ_{R,\text{eff}}}$ scales with $(F_{\text{pull}}/N_L)^2$.

\begin{figure}
	\centering
	\includegraphics[width= 0.49\textwidth]{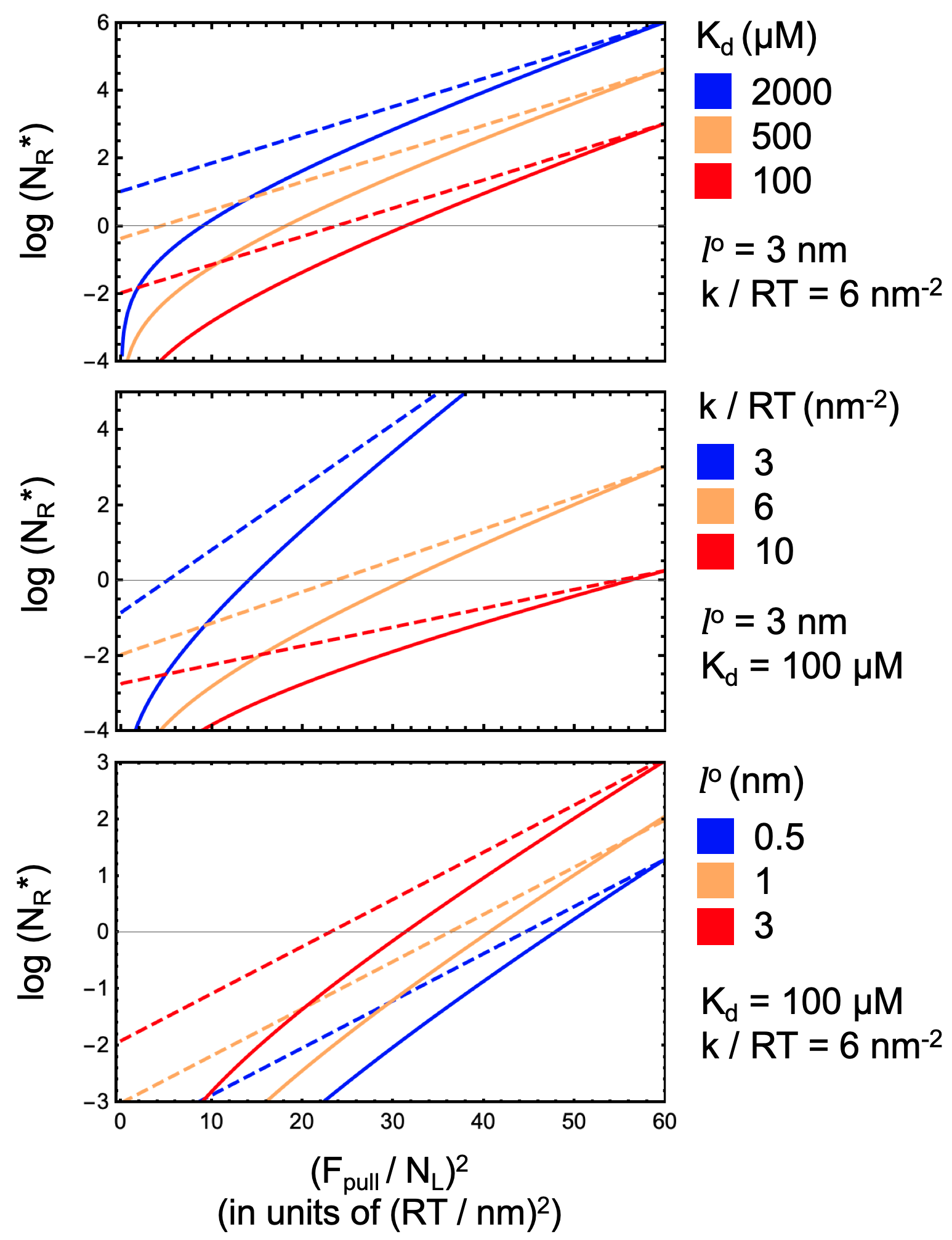}
	\caption{Plots of the logarithm of $c_R^*$ (expressed as $N_R^*$ via Eq. \ref{eqn:NREquation}) as a function of $(F_{\text{pull}} / N_L)^2$, for various choices of equilibrium ligand length $l^\circ$, ligand spring constant $k$, and ligand/receptor dissociation constant $K_d = 1 / K_{eq}$. Solid lines are numerical calculations, and dashed lines are the predicted scaling according to Eq. \ref{eqn:CRScalingForceMech}. See the caption of Figure \ref{fig:CRIntrScalings} for additional details.}
	\label{fig:CRMechScalings}
\end{figure} 

We now turn to an examination of the mechanical transition point $c_R^*$. In contrast to the intrinsic transition point,  $c_R^*$ does \emph{not} depend on the concentration $[M]$ of the particles in bulk. This is because $c_R^*$ is related only to the gradient of the free energy profile $\Delta G(h; c_R)$, whereas the chemical potential $\mu$ only shifts all $\Delta G(h; c_R)$ values by a constant.

Using Eq. \ref{eqn:CriticalForceApprox}, we can estimate the scaling of the critical receptor density $c_R^*$ given $F_{\text{pull}}$:
\begin{equation}
	\ln{c_{R}^*} \propto \ln{\left(\frac{l^\circ}{K_{\text{eq}}}\right)} + \frac{1}{2kRT} \left(\frac{F_{\text{pull}}}{N_L}\right)^2.
	\label{eqn:CRScalingForceMech}
\end{equation}
Thus, $c_R^*$ actually has the same scaling dependence as $c^\circ_{R,\text{eff}}$ on $F_{\text{pull}}$ and the other multivalent design parameters, \emph{except} for the concentration dependence. Figure \ref{fig:CRMechScalings} presents results for how $c_R^*$ varies with $F_{\text{pull}}$, for various choices of multivalent design parameters. As expected by the scaling relation above, changes in $l^\circ$ or $K_{eq}$ lead to vertical shifts in $\ln{c_R^*}$ as a function of $(F_{\text{pull}} / N_L)^2$, while changes in the ligand stiffness $k$ cause the \emph{slope} of the curve to change.

\section{Tuning the Crossover Point by Multivalent Concentration}

The fact that the intrinsic transition point, but not the mechanical transition point, depends on the particle concentration $[M]$ can be used advantageously in experimental design. To understand this, we take a deeper look at what controls the crossover between the super-selective and hyper-selective binding regimes.

Consider a fixed multivalent particle design and a given concentration $[M]$. At what choice of $F_{\text{pull}}$ does the mechanical transition point $c_R^*$ exactly meet with the intrinsic transition point $c^\circ_{R,\text{eff}}$?

When the applied force $F_{\text{pull}}$ is small, the equilibrium binding free energy of the particles to a surface with receptor density $c_R$ is $\Delta G^{\text{eff}}(c_R, F_{\text{pull}})$. The binding transition inflection point occurs at $c^\circ_{R,\text{eff}}$, where $\Delta G^{\text{eff}}(c_R, F_{\text{pull}}) = 0$. The receptor density $c_R^*$ is the smallest choice of $c_R$ where multivalent binding still occurs given the applied force; for smaller $c_R$, the applied force is stronger than the rupture force $F^*(c_R)$, and multivalent binding is prohibited. 

Thus, $c_R^*$ is the smallest choice of $c_R$ in which there is a coordinate $\tilde{h}_{eq}$ along the free energy curve $\Delta G (h; c_R^*)$ where $d \Delta G (h; c_R^*)/dh = 0$. This equilibrium free energy $\Delta G^{\text{eff}}(c_R^*, F_{\text{pull}})$ is large (positive) when the applied force is small. Increasing the applied force causes $\Delta G^{\text{eff}}(c_R^*, F_{\text{pull}}) $ to decrease towards zero, and $c_R^*$ to grow larger.

Eventually, we reach a particular choice of applied force---the crossover value $F_{\text{pull}}^{X}$---where $\Delta G^{\text{eff}}(c_R^*, F_{\text{pull}})$ ``catches up'' to $\Delta G^{\text{eff}}(c_R, F_{\text{pull}})$, i.e. $\Delta G^{\text{eff}}(c_R^*, F_{\text{pull}}) = 0$. This is precisely the choice of applied force that yields $c_R^* = c^\circ_{R,\text{eff}}$. For an applied force larger than this choice, the intrinsic threshold receptor density $c^\circ_{R,\text{eff}}$ disappears since there is no longer a choice of $c_R$ where $\Delta G^{\text{eff}}(c_R, F_{\text{pull}}) = 0$.

\begin{figure}
	\centering
	\includegraphics[width= 0.50\textwidth]{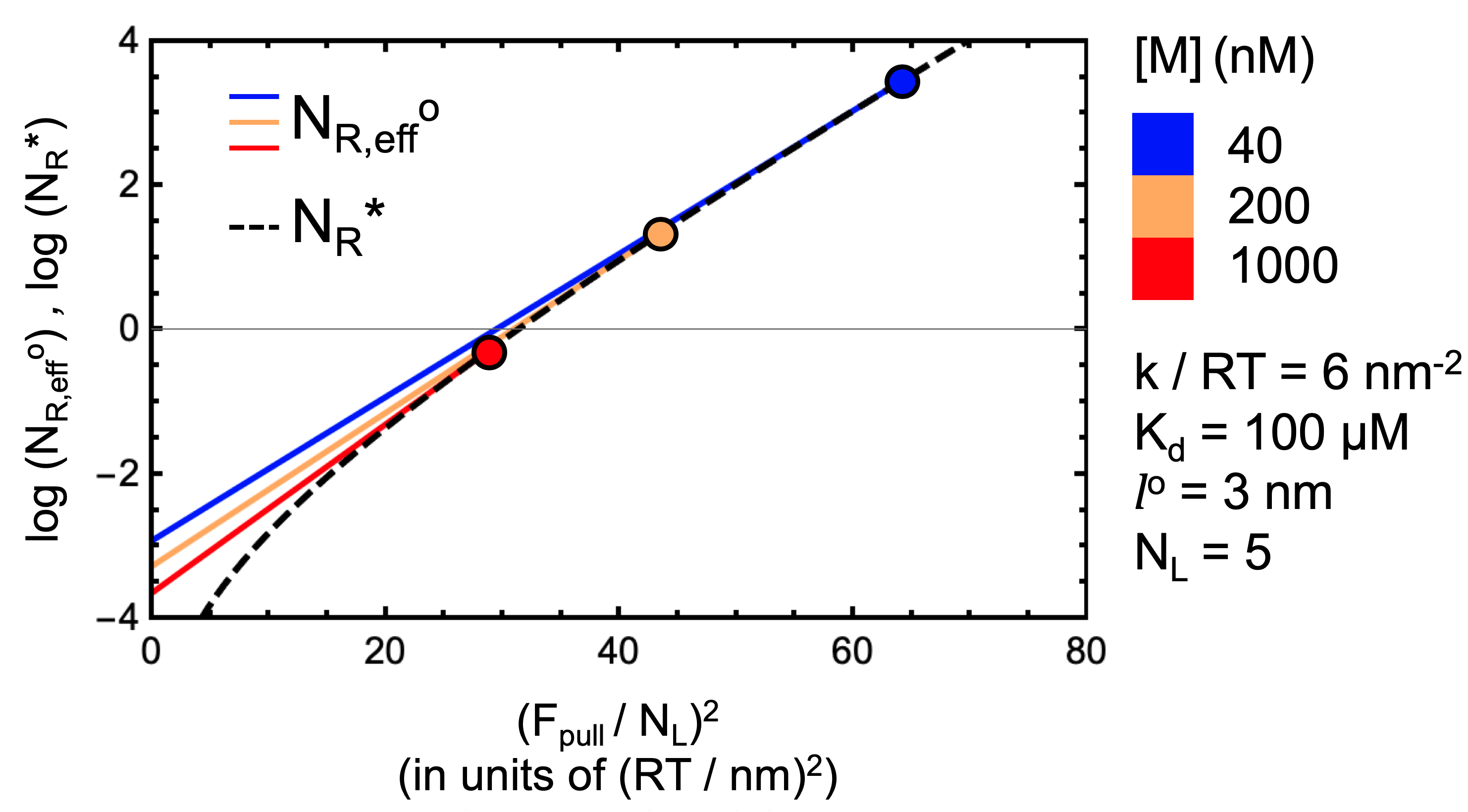}
	\caption{Plots of the logarithm of $c^\circ_{R,\text{eff}}$ (expressed as $N^\circ_{R,\text{eff}}$ via Eq. \ref{eqn:NREquation}) as a function of $(F_{\text{pull}} / N_L)^2$ for three choices of multivalent particle concentration $[M]$, given fixed particle design parameters $k$, $K_{eq}$, $l^\circ$, $N_L$. The logarithm of $c_R^*$ vs. $(F_{\text{pull}} / N_L)^2$ is also displayed for this particle design, recalling that it does not depend on the concentration $[M]$. The choice of $(F_{\text{pull}} / N_L)^2$ where $c^\circ_{R,\text{eff}} = c_R^*$ for each concentration is indicated by a coloured dot. Receptor densities are expressed as the average number of receptors within the surface footprint of a particle.}
	\label{fig:CRBothScaling}
\end{figure} 

This concept is illustrated in Figure \ref{fig:CRBothScaling}. The plot shows examples of $c_R^*$ and $c^\circ_{R,\text{eff}}$ curves, as a function of the squared force per ligand $(F_{\text{pull}} / N_L)^2$, for three choices of the multivalent particle concentration (all else being fixed). The coloured points indicate the value of applied force $F_{\text{pull}}^{X}$ where $c_R^*$ becomes equal to and then grows larger than $c^\circ_{R,\text{eff}}$. Clearly, for a fixed multivalent particle design (i.e. $k$, $K_{eq}$, $l^\circ$, $N_L$), the multivalent particle concentration determines this crossover force delineating the two binding regimes. This is a useful and simple control in experiment.

Invoking the scaling relations for $c^\circ_{R,\text{eff}}$ (Eq. \ref{eqn:CRScalingForceIntr}) and $c_R^*$ (Eq. \ref{eqn:CRScalingForceMech}) to find where $c^\circ_{R,\text{eff}} = c_R^*$ yields no dependence on $F_{\text{pull}}$, since that term has the same prefactor in both cases. If we instead suppose that the two prefactors on $(F_{\text{pull}}/N_L)^2$ differ by some amount, then we derive
\begin{equation}
	\ln{\left(\frac{c^\circ_{R,\text{eff}}}{c_R^*}\right)} = \frac{\mathcal{C}}{2 k RT} \left(\frac{F_{\text{pull}}}{N_L}\right)^2 + \frac{1}{N_L} \ln{\left(\frac{1}{[M] N_{\rm A} V_{ex}}\right)},
\end{equation}
where $\mathcal{C}$ is an unknown constant.

For diminishing choices of $F_{\text{pull}} < F_{\text{pull}}^X$, the left-hand side of this expression is positive and grows larger. Indeed, from Figures \ref{fig:CRIntrScalings}, \ref{fig:CRMechScalings}, and \ref{fig:CRBothScaling} we know that the ratio $c^\circ_{R,\text{eff}}/c_R^*$ \emph{always} gets larger with decreasing force $F_{\text{pull}}$. Thus, the constant $\mathcal{C}$ \emph{must} be negative. Pulling out the negative sign from $\mathcal{C}$ to give the positive (still unknown) constant $\mathcal{C}^+$ leads to
\begin{equation}
	\ln{\left(\frac{c^\circ_{R,\text{eff}}}{c_R^*}\right)} = -\frac{\mathcal{C}^+}{2 k RT} \left(\frac{F_{\text{pull}}}{N_L}\right)^2 + \frac{1}{N_L} \ln{\left(\frac{1}{[M] N_{\rm A} V_{ex}}\right)}.
\end{equation}

At the crossover force $F_{\text{pull}}^X$, $c^\circ_{R,\text{eff}} = c_R^*$, and so the left-hand side of this equation is zero. Solving for the value of $\left(F_{\text{pull}}^X/N_L\right)^2$ where this occurs yields
\begin{equation}
	\left(\frac{F_{\text{pull}}^X}{N_L}\right)^2 \propto \frac{2 k RT}{N_L} \ln{\left(\frac{1}{[M] N_{\rm A} V_{ex}}\right)}.
	\label{eqn:FCrossPredict}
\end{equation}
Notably, the crossover force \emph{does not have a dependence on the ligand/receptor binding constant $K_{eq}$ or ligand rest length $l^\circ$}. This can be seen in the numerical results in Figure \ref{fig:CRIntrScalings}, i.e. the values of $(F_{\text{pull}}^X/N_L)^2$ where $c^\circ_{R,\text{eff}}$ vanishes in the upper-left and bottom panels, respectively.

To understand this feature, we look back to the force-vs-extension curves in Figure \ref{fig:ForceCurves}. Notice that $K_{eq}$ and $l^\circ$ have very little influence on the \emph{slope} of the restoring force $F(h)$ as a function of displacement $\Delta z$ from the particle's equilibrium position. The only change incurred is a different value of rupture force $F^*$. Because $K_{eq}$ and $l^\circ$ do not affect the shape of the force-extension curve, then they do not have an influence on the crossover force $F_{\text{pull}}^X$. On the other hand, $F_{\text{pull}}^X$ \emph{does} depend on $k$, since $k$ affects the shape of the force-extension curves in Figure \ref{fig:ForceCurves}. The dependence of the crossover force on $k$ is seen in the middle left panel of Figure \ref{fig:CRIntrScalings}. 

Inserting Eq. \ref{eqn:FCrossPredict} back into the scaling expression for $c_R^*$, in order to estimate the crossover receptor density $c_R^X$, yields
\begin{equation}
	\ln{c_R^X} \propto \ln{\left(\frac{l^\circ}{K_{\text{eq}}}\right)} + \frac{1}{N_L} \ln{\left(\frac{1}{[M] N_{\rm A} V_{ex}}\right)}.
	\label{eqn:CRCrossPredict}
\end{equation}
Here, we now see the very clear dependence of $c_R^X$ on the ligand/receptor binding strength $K_{eq}$ and ligand length $l^\circ$, as seen in Figure \ref{fig:CRIntrScalings}. Finally, both $[M]$ and $N_L$ appear in the scaling expressions for both $F_{\text{pull}}^X$ and $c_R^X$, and indeed this is observed in the two right-hand panels in Figure \ref{fig:CRIntrScalings}. 

\section{Design rules, Experimental Considerations, Challenges}

This work has theoretically examined the adsorption thermodynamics of multivalent particles in a force field. The model consists of a solution of ligand-coated multivalent particles in contact with a flat substrate coated with point-like mobile receptors at a fixed concentration.  A given receptor may only be bound to at most one ligand at a given time, and vice-versa. The ligands themselves are modeled as harmonic springs with a given spring constant and equilibrium rest length.

The force field applies a constant force to the particles along the \emph{normal} axis of the receptor-coated substrate. Focus was placed on distinguishing between the microscopic physics that result when: the force field is applied \emph{only to bound particles}; and when the force field is applied to \emph{both} bound and unbound particles. 

For weak or no applied force, multivalent binding is super-selective and continuous with respect to the concentration of receptors on the surface. A weak applied force simply shifts the inflection point of the adsorption curve to larger values of receptor concentration.

At \emph{large} applied force, multivalent particles may only bind when the surface receptor density is larger than a critical value necessary to keep the particles anchored within the force field. When the force field is only applied to bound particles, the multivalent adsorption/desorption profile exhibits \emph{first-order} discontinuous behaviour as a function of receptor density. We refer to this adsorption behaviour as a hyper-selective binding.  However, under equilibrium conditions, the force field is placed on all particles in the system, and the adsorption profile remains a shifted continuous (second-order) transition albeit sharper.

In experiment, the multivalent particle design is often fixed by chemistry. Therefore, the most convenient variables to vary are the molar concentration $[M]$ of the multivalent particles in solution, and the applied force $F_{\text{pull}}$.
For the case where only the bound multivalent particles are susceptible to the force field, these two parameters drive the binding behaviour into one of three regimes as follows, summarised in Figure \ref{fig:BindingRegimes}:
\begin{enumerate}
	\item{At zero force $F_{\text{pull}}$, $[M]$ determines the surface receptor density $c_R^\circ$ where multivalent adsorption occurs. This is standard multivalent binding, having a continuous and super-selective binding profile $\theta(c_R)$ with an inflection point centered near $c_R^\circ$. The transition point $c_{R}^\circ$ is pushed to larger values by \emph{decreasing} the bulk multivalent particle concentration.}
	\item{Applying a non-zero force $F_{\text{pull}}$ shifts the binding transition to a new receptor density $c_{R,\text{eff}}^\circ > c_R^\circ$. This intrinsic transition point $c_{R,\text{eff}}^\circ$ grows larger by increasing the applied force, and smaller by increasing the particle concentration $[M]$. The force also defines a mechanical transition point $c_R^*$; the receptor density $c_R$ on the surface must be larger than $c_R^*$ for any binding to occur. The mechanical transition point $c_R^*$ grows larger with increasing $F_{\text{pull}}$, while it has \emph{no} dependence on the particle concentration. From Eq. \ref{eqn:FCrossPredict}, if $[M]$ and $F_{\text{pull}}$ are chosen such that
\begin{equation}
	\mathcal{F} \equiv \cfrac{F_{\text{pull}}}{\sqrt{2 k N_L RT \ln{\left(\frac{1}{[M] N_{\rm A} V_{ex}}\right)}}} \ll 1,
\end{equation}
then the force is sufficiently weak and the adsorption transition is continuous and superselective.}
	\item{On the other hand, if $[M]$ and $F_{\text{pull}}$ are chosen such that
\begin{equation}
	\mathcal{F} \gg 1\;,
\end{equation}
	then the force is strong to pull the particle away from the surface; the binding transition is likely to be hyper-selective.}
\end{enumerate}

The ligand-receptor binding constant $K_{eq}$ (among the other multivalent design parameters) influences the order of magnitude of surface receptor density where the crossover from super-selective to hyper-selective binding occurs. This was noted in the results in Figure \ref{fig:CRIntrScalings} and in the relation given by Eq. \ref{eqn:CRCrossPredict}.

If $K_{eq}$ is large, i.e. the ligands are strong-binding, then the crossover to the hyper-selective regime may occur at vanishingly small surface receptor densities. Conversely, if $K_{eq}$ is very small, then the crossover receptor density may be inaccessibly large.

From Eq. \ref{eqn:CRCrossPredict}, we can derive an estimating factor to help in diagnosing this limitation:
\begin{equation}
	\mathcal{R} \equiv \frac{l^\circ}{K_{eq} \tilde{c}_R} \left(\frac{1}{[M] N_{\rm A} V_{ex}}\right)^{1/N_L}.
	\label{eqn:RFactor}
\end{equation}
Here, $\tilde{c}_R$ is a general magnitude of the surface receptor density that is accessible in the experiment. When we choose the value of $\tilde{c}_R$ to be exactly the crossover receptor density $\mathcal{R}$ is unity. If $\mathcal{R} \gg 1$, then the input receptor concentration is well under the crossover value, while the opposite is true for $\mathcal{R} \ll 1$.

In practise, the estimating factor $\mathcal{R}$ is best be used by two calculations: once for the lowest accessible $c_R$, and another time for the largest accessible value. If the two resulting values of $\mathcal{R}$ are sufficiently greater than and less than unity, respectively, then this means that the range of receptor densities accessible in experiment are likely sufficient for catching the crossover from super-selective to hyper-selective binding when different forces are applied. 

If this is not the case, then the multivalent concentration $[M]$ is a convenient control parameter for adjusting the range of receptor densities where the crossover is expected. Indeed, Eq. \ref{eqn:RFactor} indicates how the crossover receptor density can be made larger (smaller) by decreasing (increasing) the particle concentration $[M]$ in solution.

A key finding of this study is that hyper-selective binding is obtained when a force field is applied only to bound particles. This is analogous to stating that unbound multivalent particles must be allowed to reach and bind to the surface without needing to perform thermodynamic work against the field. This poses a challenge in practise. We now outline a few experimental scenarios where hyper-selective binding might be realised. 

One possibility is to employ a continuous (slow) flow of the solution of multivalent particles parallel to the surface, so that particles pulled away by an applied force are continuously replenished by the flow. Another scenario is to use the surface-parallel flow as the source of the applied force itself. However, the statistical mechanics of multivalent binding when the applied force vector is parallel to the surface are substantially more complex than the present theory considers. For example, ligands that are highly stretched in a given state may detatch and rebind to a closer receptor. The particles will thus ``walk'' along the surface through successive ligand unbinding/rebinding events. On the other hand, when the force is normal to the adsorbing surface, stretched ligands can only relax by unbinding.

On the side of structural possibilities, we began in the Introduction at multivalent particles with inert polymers grafted to their surface. The polymers act as springs, which effectively impose a constant force on the host multivalent particle when bound. The magnitude of the force grows larger with the length of the polymers. The result is that the binding transition becomes sharper and shifts to larger values of surface receptor density, as expected by the present theory. 

This recipe as it stands cannot achieve first-order binding, as a multivalent particle must initially do work against the ``force field''---in this case, the free energy cost for compressing the inert polymers---in order to form bonds with the surface receptors. However, we can effectively turn off the force field for unbound particles in this system by designing the particles to have a ``triggered'' release of their inert polymers only when bound to the surface.

The trigger could be: an external stimulus like a change in pH or solvent composition, proximity to the surface receptors (or other surface-bound species), or binding of the sticky ligands themselves. Once triggered, the inert polymers would uncoil and expand around their host particles, effectively ``switching on'' the imposed force field for the bound particles. On the other hand, the particles would be designed so that they retract their inert polymers into a tightly-coiled configuration around the particle core when not bound to the surface. In this way, the force field is only imposed by the inert polymers once their host particle is surface-bound, and not while the particle is approaching the surface.

A more detailed treatment of multivalent force response would also take into account thermal fluctuations of the multivalent particles along their free energy landscapes (e.g. in Figure \ref{fig:FE}). For large particles with many ligand/receptor bonds, the free energy landscapes will be quite deep, and thermal fluctuations will play a minimal role. However for small multivalent binders with shallow binding free energy profiles, fluctuations will tend to blur the sharpness of the hyper-selective regime.

The majority of the discussion has focused on multivalent \emph{adsorption}. However, the binding can be equivalently examined from the perspective of force-induced \emph{desorption}. The present theory may be useful as a starting point for predicting which multivalent particles will remain bound, under an applied force, on a surface with a heterogeneous distribution of fixed receptors. For example, if the particles have a magnetic dipole, then activating a magnetic field gradient will impart a pulling force on the dipoles. Particles will only remain bound where the local receptor density is high enough, i.e. the local rupture force is larger than the applied force. This theory can be used to predict the necessary threshold receptor density required for survival.

\section{Acknowledgments}

This work has been carried out whilst financially supported by the Netherlands 4TU.High-Tech Materials research programme `New Horizons in designer materials' (www.4tu.nl/htm). We wish to thank Lorenzo Albertazzi for great discussion and inspiration on the topic of this work, as well as Daan Frenkel and Stefano Angioletti-Uberti for feedback on the manuscript.

\appendix

\section{Derivation of the multivalent force-responsive model}
\label{app:Theory}

Consider a multivalent particle with $N_L$ ligands that interact with mobile receptors on an adjacent flat surface. The density of receptors on the surface is $c_R$ (in units of moles of receptors per squared length $b$), and we assume that the receptors cannot be depleted. Let $z$ be the coordinate axis extending orthogonal to the receptor substrate. Along this axis, we define $h$ to be the distance between the receptor surface, and the surface of the multivalent particle to which the ligands are tethered.

The ligands are treated as Hookian springs with a spring contant of $k$ and rest length of $l^\circ$, while the receptors are defined to be mobile points on the substrate. The ligands have an individual force-extension equation of
\begin{equation}
	F_{\text{lig},\text{spring}}(z) =  -k \left(z - l^\circ\right)
\end{equation}
where $z$ is their extension length, $R$ is the ideal gas constant, and $T$ is temperature. Thus, the contribution from this term to the free energy of a ligand when the substrate and multivalent particle surface are separated by a gap of size $h$ is
\begin{equation}
	\Delta G^b_{\text{lig},\text{spring}}(h) = \frac{1}{2} k (h - l^\circ)^2.
	\label{eqn:AppLigStretchFreeEnergy}
\end{equation}
Ligand/receptor bonding is the second contribution to the free energy of a ligand. From equilibrium multivalency theory \cite{Varilly:2012gl}, this takes the form
\begin{equation}
	\frac{\Delta G^b_{\text{lig},\text{bonding}}(h)}{RT} =  - \ln{\left(K_{\text{eq}} [C]_{\text{eff}}(h)\right)}.
\end{equation}
where $K_{\text{eq}}$ is the ligand/receptor equilibrium constant in free solution (in units of inverse molarity), and $[C]_{\text{eff}} (h)$ is the effective molarity of the receptors. The effective molarity is calculated by
\begin{equation}
	[C]_{\text{eff}}(h) = \frac{c_R}{h}.
\end{equation}
Thus, the free energy of each bound ligand is given by
\begin{align}
	\frac{\Delta G^b_{\text{lig}}(h)}{RT} &= \frac{\Delta G^b_{\text{lig},\text{spring}}(h)}{RT} + \frac{\Delta G^b_{\text{lig},\text{bonding}}(h)}{RT}\nonumber \\
	&= \frac{k (h - l^\circ)^2}{2 RT} - \ln{\left(\frac{c_R K_{\text{eq}}}{h}\right)} \label{eqn:LigandQPartway}.
\end{align}
The full equilibrium partition function for the multivalent particle when bound to the receptor surface is thus
\begin{equation}
	Q_{b}(h) = \left(1 + e^{-\Delta G^b_{\text{lig}}(h)/RT}\right)^{N_L}.
\end{equation}
This partition function represents the fact that each of the $N_L$ ligands on the surface can be independently bound or unbound to a receptor. The binding free energy of the whole multivalent particle is therefore
\begin{equation}
	\frac{\Delta G_{b}(h)}{RT} = -\ln{Q_{b}(h)} = -N_L \ln{\left(1 + e^{-\Delta G^b_{\text{lig}}(h)/RT}\right)}.
	\label{eqn:AppFreeEnergyFullPartWay}
\end{equation}
The external force required to push the particle to some displacement $h$ is the gradient of this free energy of binding as a function of displacement height $h$:
\begin{align}
	&\frac{F(h)}{RT} = \frac{d (\Delta G_{b}(h)/RT)}{d h} \nonumber \\
	&= N_L \left(\frac{e^{-\Delta G^b_{\text{lig}}(h)/RT}}{1 + e^{-\Delta G^b_{\text{lig}}(h)/RT}}\right) \left[\frac{k (h - l^\circ)}{RT} + \frac{1}{h}\right],
	\label{eqn:AppForceBad}
\end{align}
A point of concern here is that as $h$ approaches $0$---i.e. as we push the particle towards the adsorbing surface---the effective molarity contribution to the ligand binding free energy in Eq. \ref{eqn:LigandQPartway} diverges to infinity. This will never be out-competed by the ligand stretch free energy, Eq. \ref{eqn:AppLigStretchFreeEnergy}. The result is that $\Delta G_{b}(h)$ grows infinitely deep at $h = 0$, which is not physical.

\begin{figure*}
	\centering
	\includegraphics[width= 0.9\textwidth]{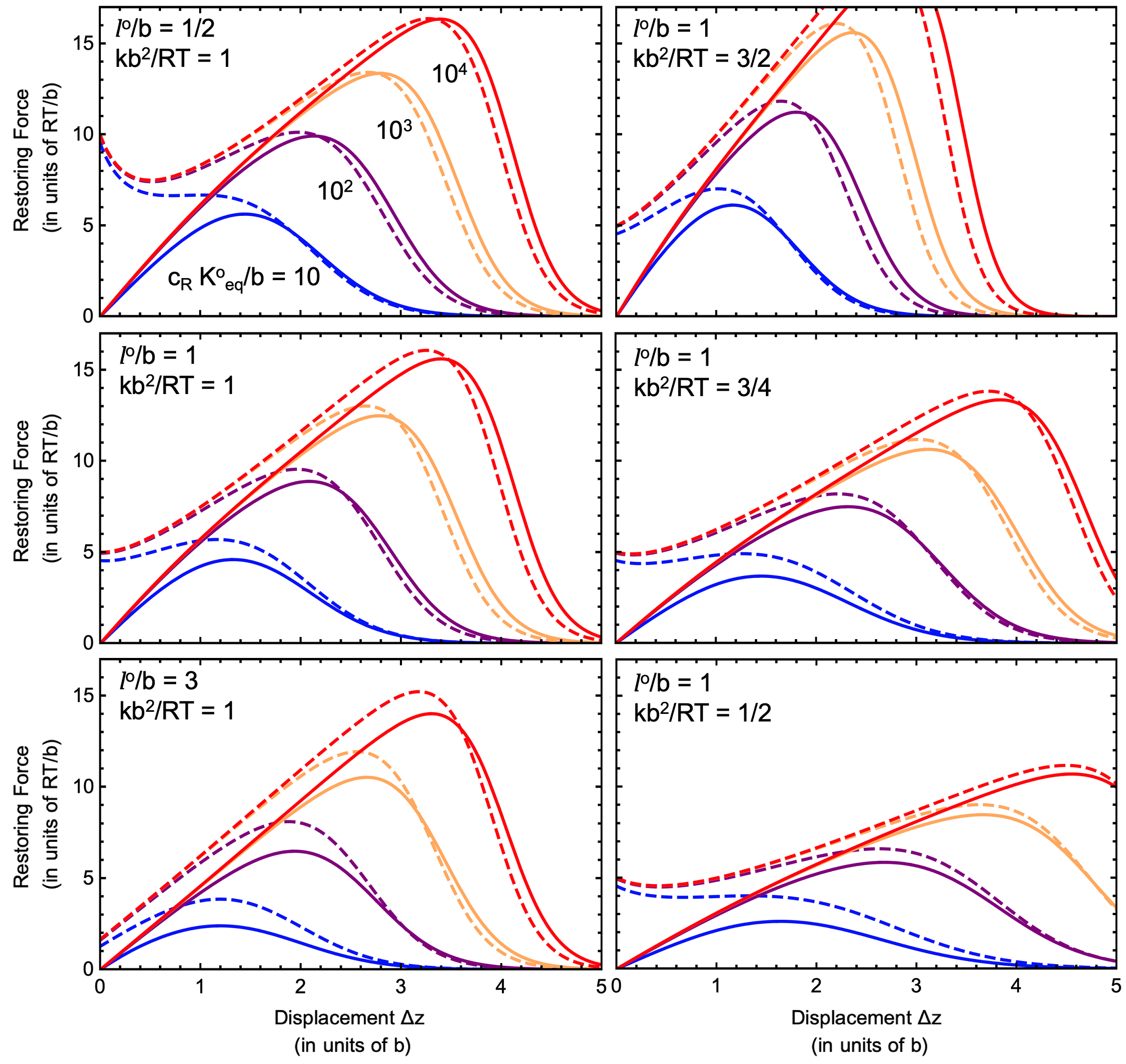}
	\caption{Plots of the restoring force with (Eq. \ref{eqn:AppForce}, solid lines) and without (Eq. \ref{eqn:AppForceBad}, dashed lines) ligand repulsion, as a function of relative separation distance $\Delta z$ between the multivalent particle and the receptor surface. Each panel shows results for four choices of the effective ligand-receptor binding strength, $c_R K_{\text{eq}}/b$. Results for different choices of initial equilibrium binding height $h_{bind}/b$, and ligand length $N$, are also given. All plots consider multivalent particles with $N_L = 5$ ligands.}
	\label{fig:AppForceCurvesBad}
\end{figure*}  

The origin of this problem is that Eq. \ref{eqn:AppFreeEnergyFullPartWay} is missing a repulsive free energy contribution for the loss of ligand configurational entropy, which grows substantial when $h$ becomes small compared to $l^\circ$. To combat this, we must implement an additional potential. 

We define this potential to be the entropic penalty for the ligands to be confined within the space $h$ between the substrate and multivalent particle exterior.\cite{Varilly:2012gl} There are three scenarios to be considered. The first is what we define as the reference state of a ligand: when it is unbound given that the multivalent particle is at infinite distance from the surface (i.e. $h = \infty$). The next case is when the ligand is unbound and the multivalent particle is positioned at $h$, and the last case is when the ligand is bound to a receptor and the host particle is at distance $h$. We now consider these three scenarios in turn.

A ligand is treated as two equally-sized pieces, each with rest length $l^\circ / 2$ and a spring constant of $k_{\text{sub}} = 2 k$ (the latter following from Hookian springs in series). One subsection is attached to the multivalent particle, and the other is imagined to be attached to a receptor on the substrate. (This is equivalent to taking the perspective where the receptors are now flexible entities with rest length $l^\circ / 2$ and spring constant $k_{\text{sub}}$.) The two remaining ends of the subsections are dangling, referred to as ``binding tips''. 

The probability distribution for the binding tip of one ligand subsection is
\begin{align}
	P_{\text{tip}}(z) &= \sqrt{\frac{k_{\text{sub}}}{2 \pi RT}} \exp{\left(-\frac{k_{\text{sub}} (z - \frac{l^\circ}{2})^2}{2 RT}\right)} \nonumber \\
	&= \sqrt{\frac{k}{\pi RT}} \exp{\left(-\frac{k (z - \frac{l^\circ}{2})^2}{RT}\right)}
\end{align}
For mathematical simplicity, we restrict the configurational freedom of the binding tips to only lie in the $z$ axis. When the multivalent particle is infinitely far from the receptor surface, then the configurational integral for the two subsections of a ligand is
\begin{align}
	Z_{ub}^\circ &= \left\{\int_{0}^{\infty}{P_{\text{tip}}(z) \ dz}\right\}^2 \nonumber \\
	&= \left\{\frac{1}{2} \left[\erf{\left(\frac{l^\circ}{2} \sqrt{\frac{k}{RT}}\right)} + 1\right]\right\}^2.
\end{align}
The integral is squared because both of the ligand subsections are configurationally independent of each other. Obviously $Z_{ub}^\circ$ is just the reference state, and so it has no dependence on the particle-surface separation $h$. Note that the error function $\erf{(x)}$ is defined in the standard way to be
\begin{equation}
	\erf{(x)} = \sqrt{\frac{4}{\pi}} \int_{0}^{x}{e^{-t^2} \ dt}.
\end{equation}
Next, when the multivalent particle is at $h$, then the binding tips of the ligand subsections must reside between between $z = 0$ and $z = h$:
\begin{align}
	&Z_{ub}(h) = \left(\int_{0}^{h}{P_{\text{tip}}(z) \ dz}\right)^2 \nonumber \\
	&= \left\{\frac{1}{2} \left[\erf{\left(\frac{l^\circ}{2}\sqrt{\frac{k}{RT}}\right)} + \erf{\left( \left(h - \frac{l^\circ}{2}\right) \sqrt{\frac{k}{RT}}\right)}\right]\right\}^2.
\end{align}
This leads us to the definition of the configurational free energy for an unbound ligand, relative to when the multivalent particle is at infinite distance from the surface:
\begin{equation}
	\frac{\Delta G^{ub}_{\text{lig},\text{cnf}}(h)}{RT} \equiv -\ln{\left(\frac{Z_{ub}(h)}{Z_{ub}^\circ}\right)}.
\end{equation}
When a ligand is bound to a receptor, then this is equivalent to when the binding tips of the two ligand subsections are constrained to lie at the same coordinate $z$:
\begin{align}
	Z_{b}(h) &= \int_{0}^{h}{\Delta w \times P_{\text{tip}}(z) P_{\text{tip}}(h - z) \ dz} \nonumber \\
	&= \exp{\left(-\frac{k (h - l^\circ)^2}{2RT}\right)} \times \sqrt{\frac{k \Delta w^2}{2 \pi RT}} \erf{\left(\sqrt{\frac{k h^2}{2 RT}}\right)}
\end{align}
where $\Delta w$ is the necessary distance (in units of $b$) between the two binding tips for them to be considered ``bound'', taking the role of a ``localisation length''. We will assume that this is unity throughout.

We see that this approach of dividing the ligands into two subsections naturally yields our original spring term for the full ligand, $\exp{[-k (h - l^\circ)^2 / 2RT]}$, which was placed into the bound ligand partition function in Eq. \ref{eqn:AppLigStretchFreeEnergy}. The error function then properly accounts for the configurational space of the ligand within the gap $h$ between the receptor substrate and multivalent particle surface. Maintaining our definition for $\Delta G^b_{\text{lig},\text{spring}}(h)$ as above, then the configurational entropy of a bound ligand is defined to be just the residual part in $Z_{b}(h)$ not contained in $\Delta G^b_{\text{lig},\text{spring}}(h)$:
\begin{equation}
	\frac{\Delta G^{b}_{\text{lig},\text{cnf}}(h)}{RT} \equiv -\ln{\left(\frac{Z_{b}(h)}{Z_{ub}^\circ}\right)} - \frac{\Delta G^{b}_{\text{lig},\text{spring}}(h)}{RT}.
\end{equation} 
The repulsive ligand potentials $\Delta G^b_{\text{lig},\text{cnf}}$ and $\Delta G^{ub}_{\text{lig},\text{cnf}}$ can now be incorporated into the total ligand free energies, so that they read
\begin{widetext}
\begin{align}
	\frac{\Delta G^b_{\text{lig}}(h)}{RT} &= \frac{\Delta G^b_{\text{lig},\text{spring}}(h)}{RT} + \frac{\Delta G_{\text{lig},\text{bonding}}(h)}{RT} + \frac{\Delta G^b_{\text{lig},\text{cnf}}(h)}{RT}\nonumber \\
	&=  \frac{k (h - l^\circ)^2}{2 RT} - \ln{\left(\frac{c_R K_{\text{eq}}}{h}\right)} -\ln{\left(\sqrt{\frac{k \Delta w^2 }{2 \pi RT}} \erf{\left(\sqrt{\frac{k h^2}{2 RT}}\right)}\right)} + \ln{Z_{ub}^\circ} \label{eqn:LigandQ} \\
	\frac{\Delta G^{ub}_{\text{lig}}(h)}{RT} &= \frac{\Delta G^{ub}_{\text{lig},\text{cnf}}(h)}{RT} = -2 \ln{\left\{\frac{1}{2} \left[\erf{\left(\frac{l^\circ}{2}\sqrt{\frac{k}{RT}}\right)} + \erf{\left( \left(h - \frac{l^\circ}{2}\right) \sqrt{\frac{k}{RT}}\right)}\right]\right\}} + \ln{Z_{ub}^\circ} \label{eqn:LigandUBQ}.
\end{align}
\end{widetext}
This leads to the multivalent particle binding free energy analogous to Eq. \ref{eqn:AppFreeEnergyFullPartWay}:
\begin{align}
	&\frac{\Delta G_{b}(h)}{RT} = -\ln{Q_{b}(h)} \nonumber \\
	&= -N_L \ln{\left(e^{-\Delta G^{ub}_{\text{lig}}(h)/RT} + e^{-\Delta G^b_{\text{lig}}(h)/RT} \right)}.
	\label{eqn:AppFreeEnergyFull}
\end{align}
The equation for the restoring force of the multivalent particle is obtained by
\begin{align}
	&\frac{F(h)}{RT} = \frac{d}{dh} \left(\frac{\Delta G_{b}(h)}{RT}\right) \nonumber \\
	&= N_L \times \left[P_{b,1}(h) F_b(h) - (1 - P_{b,1}(h)) F_{ub}(h)\right].
	\label{eqn:AppForce}
\end{align}
Here, the single-ligand binding probability $P_{b,1}(h)$ is given by
\begin{equation}
	P_{b,1}(h) = \frac{e^{-\Delta G^b_{\text{lig}}(h)/RT} }{e^{-\Delta G^{ub}_{\text{lig}}(h)/RT}  + e^{-\Delta G^b_{\text{lig}}(h)/RT} },
	\label{eqn:PRecBound}
\end{equation}
and the two contributions to the force are
\begin{align}
	&F_{b}(h) = \frac{k (h - l^\circ)}{RT} + \frac{1}{h} - \dfrac{\exp{\left(- \frac{k h^2}{2 R T}\right)} \sqrt{\frac{2 k \Delta w^2}{\pi R T}}}{\erf{\left(\sqrt{\frac{k h^2}{2 R T}}\right)}} \\
	&F_{ub}(h) = \sqrt{\frac{4k}{\pi R T}} \nonumber \\
	& \times \exp{\left[-\left(\frac{k (h - \frac{l^\circ}{2})^2}{R T} -\frac{\Delta G^{ub}_{\text{lig}}(h) - RT \ln{Z_{ub}^\circ}}{2 RT}   \right)\right]}.
\end{align}
Figure \ref{fig:AppForceCurvesBad} shows examples of multivalent force-extension curves with and without the repulsive ligand potentials. For ease of comparison, we plot results as a function of the relative displacement variable $\Delta z$. For the calculations without the ligand repulsion terms, $\Delta z \equiv h - l^\circ$; for those with the repulsion terms, we define $\Delta z \equiv h - h_{\text{min}}$, where $h_{\text{min}}$ is the height coordinate that minimizes the binding free energy $\Delta G_{b}/RT$ (i.e. Eq. \ref{eqn:AppFreeEnergyFull}). 

The ligand repulsion free energy terms adjust the behaviour of the force-extension curves at low displacements $\Delta z$, so that they don't unphysically diverge as the multivalent particle draws next to the receptor surface. (This is particularly notable in the upper-left panel of Figure \ref{fig:AppForceCurvesBad}.) However, the ligand repulsion terms have little influence on the relevant portion of the force-extension curve, near the rupture point. Comparing the dashed and solid curves in Figure \ref{fig:AppForceCurvesBad} reveals that both the magnitudes $F^*$ and coordinates $h^*$ for rupture change little, except when particle binding is very weak.

\section{Approximate equation for the rupture force $F^*$}
\label{app:Approximations}

In this section we examine the scaling behaviour of $F^*$ for large overall ligand binding strength $c_R K_{\text{eq}}$. The condition for the binding height $h^*$ where rupture occurs, i.e. $d F(h) / d h = 0$, cannot be obtained analytically in general. To make progress, we make the following assumptions and approximations:
\begin{enumerate}
	\item{Dissociation of the multivalent particle occurs when the probability that a single ligand is bound, $P_{1,b}(h)$, decreases to a critical value $P_{1,b}(h)^*$ that is independent of the input parameters for the system.}
	\item{The dissociation distance $h^*$ is sufficiently large that the ligand repulsion terms, $\Delta G^{b}_{\text{lig},\text{cnf}}$ and $\Delta G^{ub}_{\text{lig},\text{cnf}}$, are nearly zero}
\end{enumerate}
Under these approximations, then
\begin{equation}
	P_{1,b}(h) \approx \frac{q_b(h)}{1 + q_b(h)}
\end{equation}
where
\begin{equation}
	q_b(h) = \frac{c_R K_{\text{eq}} e^{-\frac{k (h - l^\circ)^2}{2 RT}}}{h}.
\end{equation}
Thus, choosing a critical value of $P_{1,b}(h)$ implies choosing a critical value of the quantity $q_b(h)$. Let this be called $q^*_b(h)$, and the corresponding value of $h$ where this is reached $h^*$. The expression for $q_b(h)$ can be rewritten to 
\begin{equation}
	\frac{q_b^* h^*}{l^\circ} = \frac{c_R K_{\text{eq}} e^{-\frac{k (h^* - l^\circ)^2}{2 RT}}}{l^\circ}
\end{equation}
The left-hand side can now be considered a scaled threshold value of $\tilde{q}^*_b = q_b^* h^* /l^\circ$. This equation can be solve explicitly for $h^*$ appearing on the RHS to yield
\begin{align}
	h^* &= l^\circ + \sqrt{\frac{2 RT}{k} \ln{\left(\frac{c_R K_{\text{eq}}}{l^\circ \cdot \tilde{q}_b^*}\right)}}
	\label{eqn:AppApproxZStart}
\end{align} 

\begin{figure}
	\centering
	\includegraphics[width= 0.5\textwidth]{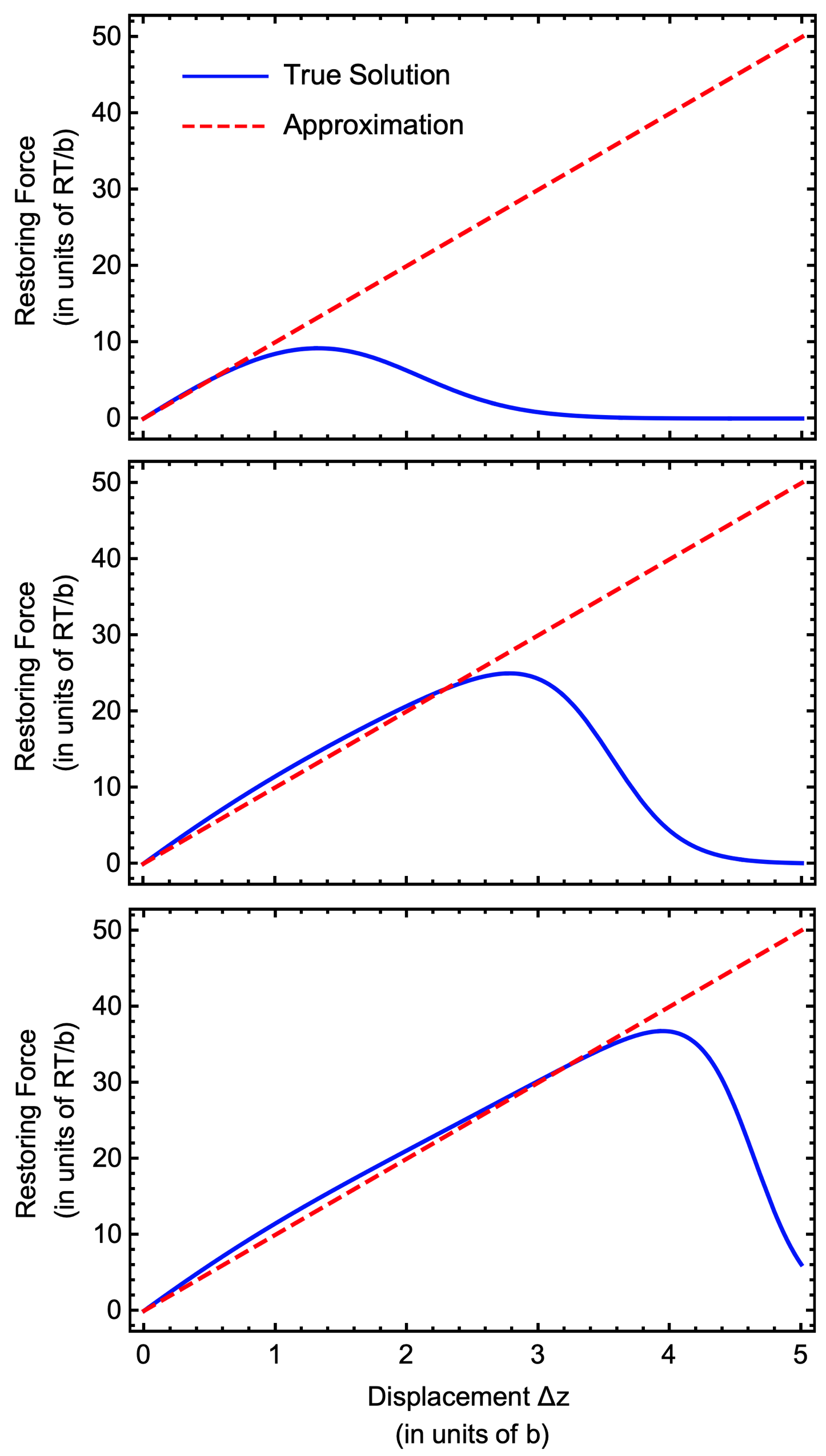}
	\caption{Plots of the true (Eq. \ref{eqn:AppForce}) and approximate (Eq. \ref{eqn:ForceApprox}) restoring force as a function of relative separation distance $\Delta z$ between a multivalent particle and the receptor surface. For the ``true'' data, this distance quantity is defined as $\Delta z = h - h_{\text{min}}$, where $h_{\text{min}}$ is the height coordinate that minimizes the binding free energy $\Delta G_{b}(h)/RT$ (i.e. Eq. \ref{eqn:AppFreeEnergyFull}). For the ``approximate'' results, $\Delta z = h - l^\circ$. From top to bottom, effective ligand-receptor binding strengths are $c_R K_{\text{eq}}/b = 10$, $10^3$, and $10^5$. All plots consider multivalent particles with $N_L = 10$ ligands each with a spring constant of $k b^2/RT = 1$.}
	\label{fig:ForceCurvesApprox}
\end{figure}

\begin{figure}
	\centering
	\subfigure[]{\includegraphics[width= 0.51\textwidth]{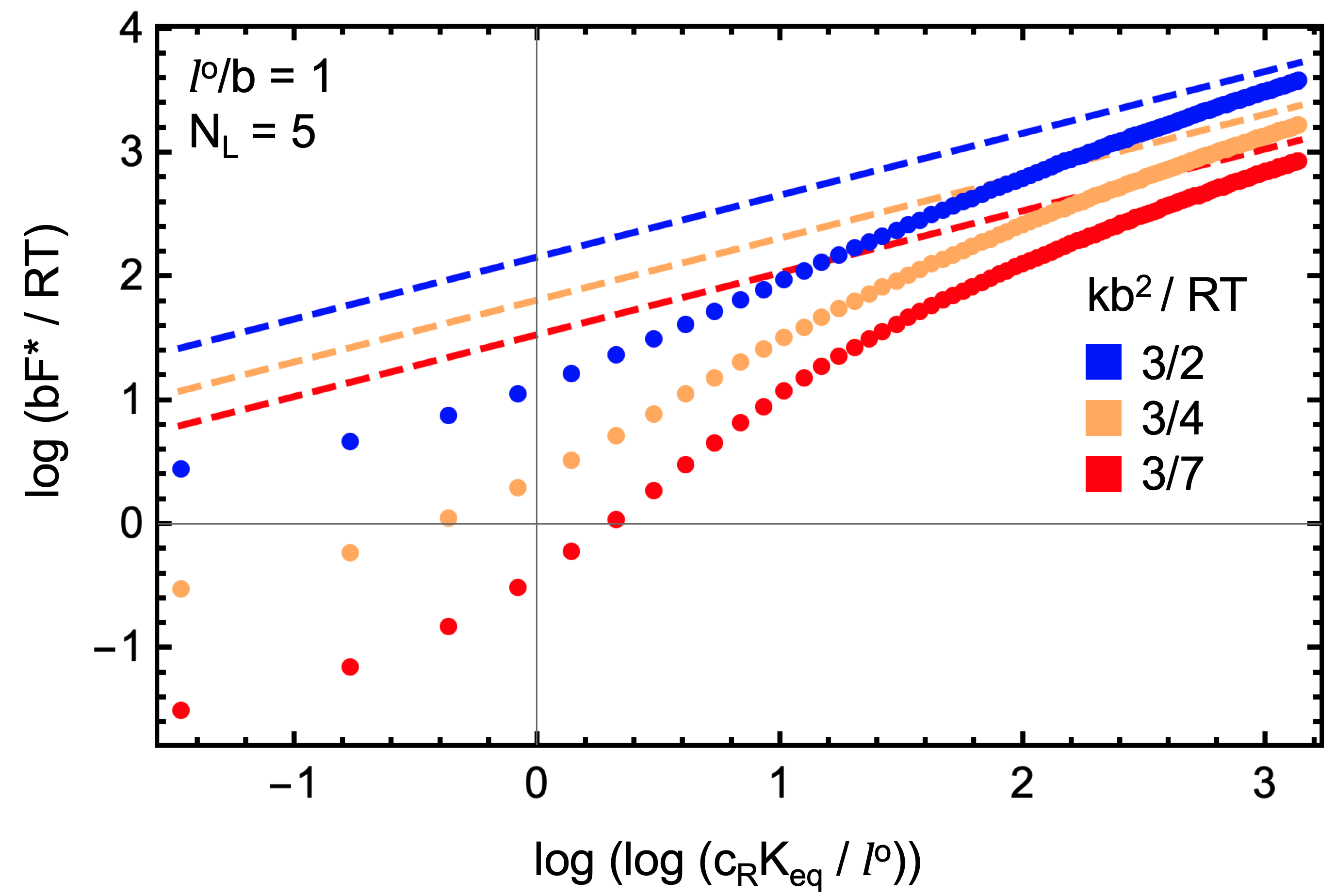}}
	\subfigure[]{\includegraphics[width= 0.51\textwidth]{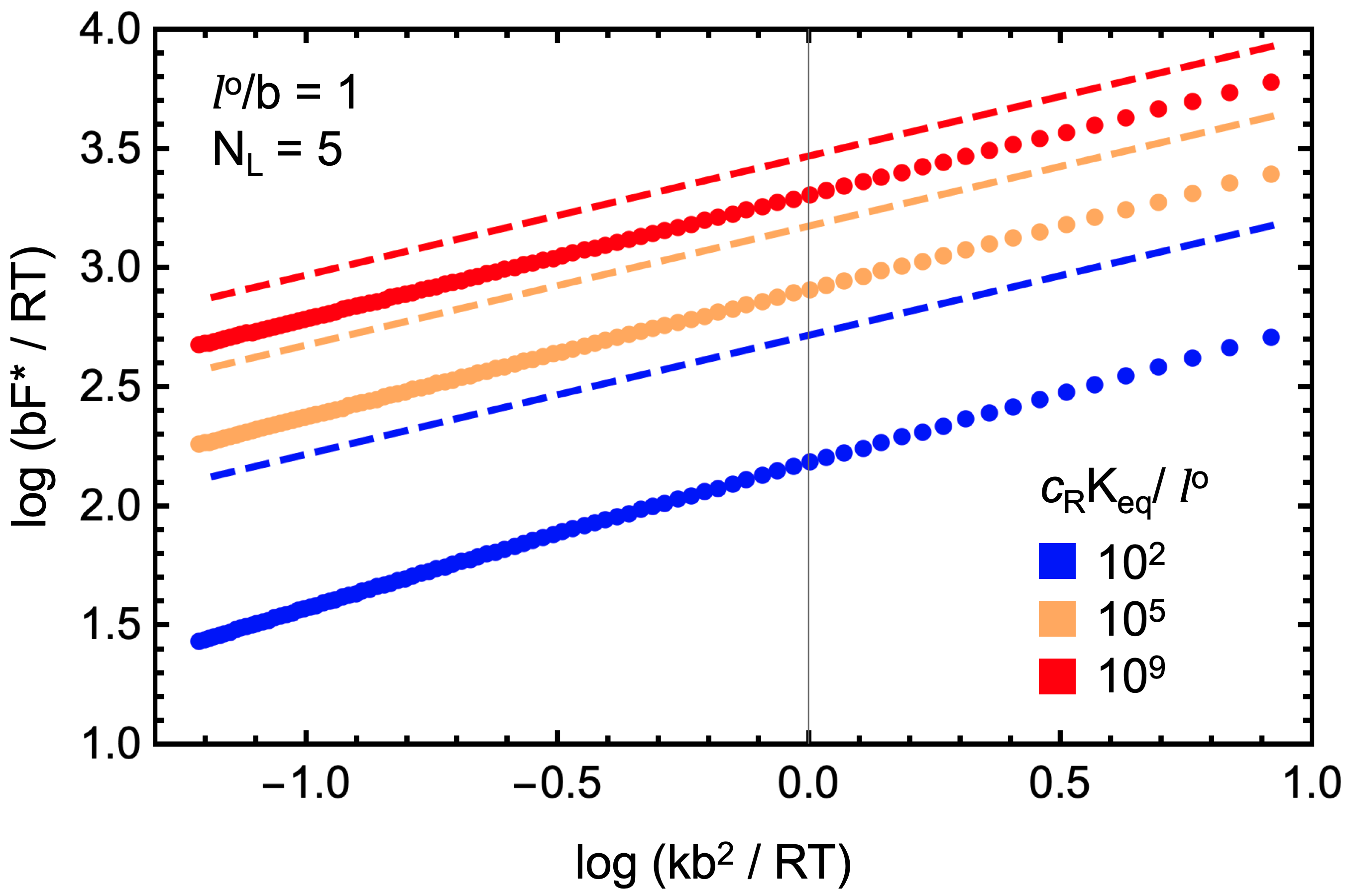}}
	\caption{Logarithm of the true (solid points) and approximate (dashed lines) rupture force $b F^*/RT$, as a function of: (a) the double logarithm of $c_R K_{\text{eq}}/l^\circ$; (b) the log of the ligand stiffness constant $k b^2/RT$ (b). The true results are obtained from the full theory, by numerically maximising Eq. \ref{eqn:AppForce}, while the approximations are computed analytically by Eq. \ref{eqn:AppBApproxA} (in a) and \ref{eqn:AppBApproxB} (in b). Parameters for each set of calculations are given within the plots.}
	\label{fig:RuptForces}
\end{figure}

To calculate the rupture force $F^*$ that this $h^*$ corresponds to, we note that for values of $h$ before $h^*$, Eq. \ref{eqn:AppForce} is well-described for increasingly large $c_R K_{\text{eq}}$ by
\begin{equation}
	\frac{F(h)}{RT} \approx N_L \left(\frac{k (h - l^\circ)}{RT}\right) = N_L \left(\frac{k \Delta z}{RT}\right).
	\label{eqn:ForceApprox}
\end{equation}
This is demonstrated in Figure \ref{fig:ForceCurvesApprox}. Inserting the approximation for $h^*$ (Eq. \ref{eqn:AppApproxZStart}) into this expression yields
\begin{equation}
	\frac{F^*}{RT} \approx N_L \sqrt{\frac{2 k}{RT} \ln{\left(\frac{c_R K_{\text{eq}}}{l^\circ \cdot \tilde{q}_b^*}\right)}}
\end{equation}
This result is compared to the true rupture forces in Figure \ref{fig:RuptForces}. We see that the approximate form properly captures the scaling of $F^*$ with $k$ (panel b)
\begin{equation}
	\ln{\left(\frac{b F^*}{RT}\right)} \propto \frac{1}{2} \ln{\left(\frac{k b^2}{RT}\right)},
	\label{eqn:AppBApproxA}
\end{equation}
over all ranges of those variables, for both large and small values of the effective ligand/receptor binding strength $c_R K_{\text{eq}}/l^\circ$. Proper scaling of the approximate equation with $c_R K_{\text{eq}}/l^\circ$ in panel (a),
\begin{equation}
	\ln{\left(\frac{bF^*}{RT}\right)} \propto \ln{\left[\ln{\left(\frac{c_R K_{\text{eq}}}{l^\circ}\right)}\right]}
	\label{eqn:AppBApproxB}
\end{equation}
is only reached when $c_R K_{\text{eq}}/l^\circ$ grows large.

\section{Verifying Bell force response}
\label{app:BellComparison}

The Bell model \cite{Bell:1978hj} describes the failure rate of individual ligand-receptor bonds in multivalent interactions, as a function of the force $F$ pulling the two host objects apart. When the two multivalent objects are connected by $N_b$ bonds, then the failure rate scales as
\begin{equation}
	\text{failure rate} \propto e^{\gamma F / N_b kT}
	\label{eqn:BellModel}
\end{equation}
where $\gamma$ is a force-response parameter. The failure rate grows exponentially with the applied force $F/N_b$ per bond.

In our present model, the characteristic failure time of an individual bond is the average time that it takes for the ligand to become unbound, counting from the time when it first formed the bond. As our model does not explicitly consider dynamics, the failure time may approximated as the reciprocal of the probability that a ligand is unbound. This is equivalent to saying that the ensemble-averaged failure rate is measured by the probability that a ligand is unbound at any given time, given that it was bound beforehand.

Mathematically, this is defined as
\begin{equation}
	\text{failure rate} \approx 1 - P_{b,1}(f).
\end{equation}
Here, $P_{b,1}(f)$ is the probability that a given ligand is \emph{bound}, and $f = F/N_L$ is the pulling force per ligand. Equation \ref{eqn:PBoundOne} expresses $P_{b,1}$ as a function of the displacement height $h$ of the multivalent particle, though via Eq. \ref{eqn:ForceReal} this quantity may be calculated as a function of the applied total force $F$ on the particle. 

\begin{figure}
	\centering
	\includegraphics[width= 0.51\textwidth]{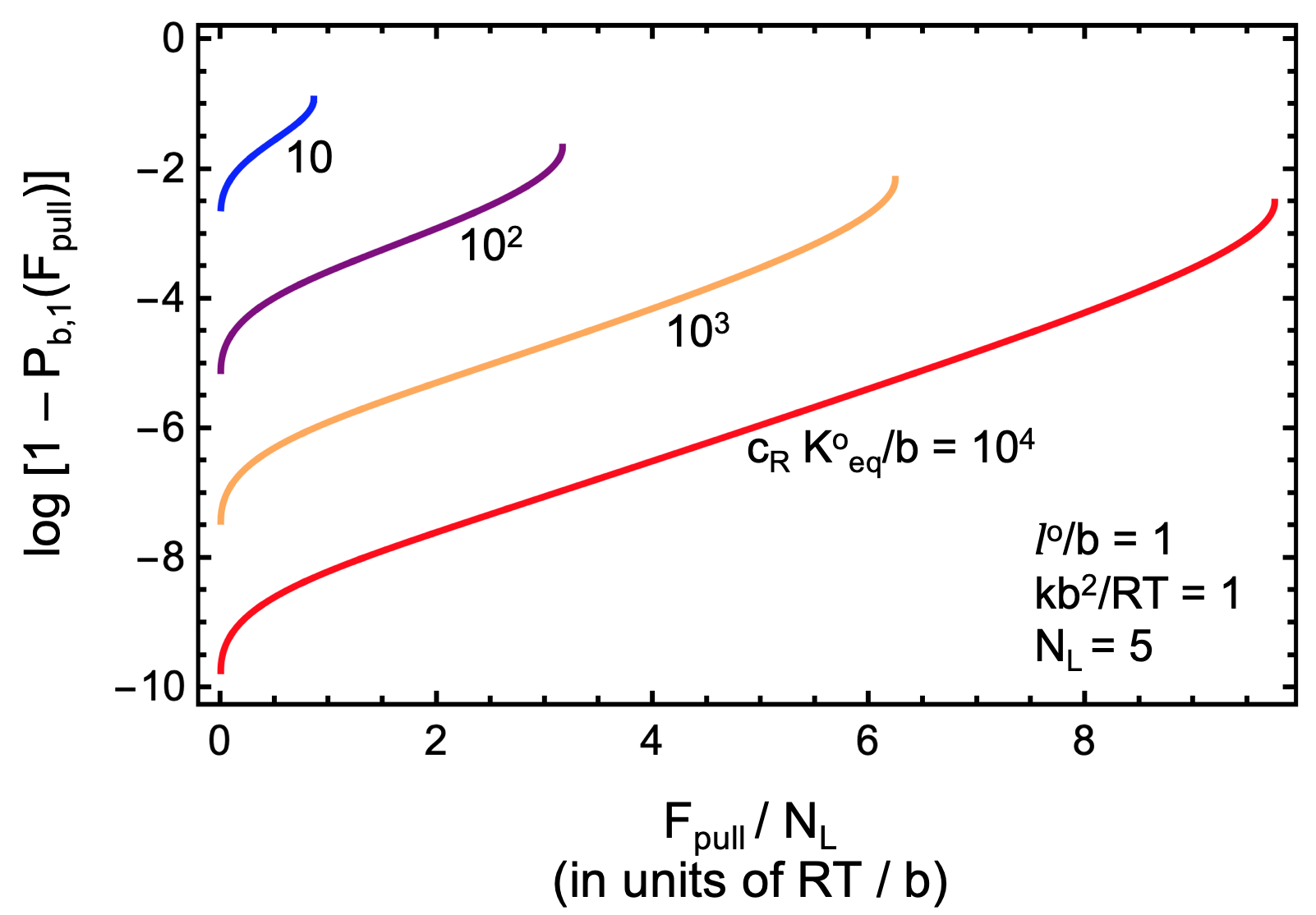}
	\caption{Logarithm of the probability that a ligand is unbound, as a function of the pulling force per ligand. Results are shown for four choices of the effective ligand-receptor binding constant $c_R K_{\text{eq}}/l^\circ$; additional parameters are given in the figure.}
	\label{fig:BellBehaviour}
\end{figure}  

If our model is to exhibit Bell behaviour, then we should expect
\begin{equation}
	\ln{\left[1 - P_{b,1}\left(f \right)\right]} \propto f
\end{equation}
according to Eq. \ref{eqn:BellModel}. Figure \ref{fig:BellBehaviour} presents plots of the left- and right-hand side quantities in this expression, for weak to strong binding ligands. The figure makes apparent, particularly for strong-binding ligands, the linear dependence between $\ln{(1 - P_{b,1}\left(f \right))}$ and the force per ligand $f$. Our model therefore behaves in accord with the Bell theory.

\bibliography{main}

\end{document}